\documentclass[aps,prx,groupedaddress,twocolumn,showpacs,longbibliography,10pt]{revtex4-1}
\pdfoutput=1
\usepackage[table]{xcolor}
\usepackage{amsmath}
\usepackage{amsthm, amssymb,bbold,slashed}    
\usepackage{braket}              
\usepackage{graphicx}
\usepackage{draft}
\usepackage{color}
\usepackage{times, verbatim}      
\usepackage{mathtools,tikz}
\usepackage{pgfplots}
\usetikzlibrary{decorations.pathmorphing}
\usetikzlibrary{matrix}
\usepackage{natbib}
\usepackage{multirow}
\usepackage{hyperref}
\usepackage{subcaption}

\newcommand{\dom}{\operatorname{dom}}
\definecolor{colour1}{rgb}{0.368417, 0.506779, 0.709798}
\definecolor{colour2}{rgb}{0.880722, 0.611041, 0.142051}
\definecolor{colour3}{rgb}{0,1,1}
\definecolor{colour4}{rgb}{0,1,0}
\definecolor{colour5}{rgb}{1,1,0}
\usetikzlibrary{decorations.pathreplacing}
\usepackage{dsfont}
\def\id{\mathds{1}}
\usepackage{amsthm}

\hypersetup{
       colorlinks=true,
}

\usepackage{multirow}
\makeatletter
\newsavebox{\@brx}
\newcommand{\llangle}[1][]{\savebox{\@brx}{\(\m@th{#1\langle}\)}%
  \mathopen{\copy\@brx\kern-0.5\wd\@brx\usebox{\@brx}}}
\newcommand{\rrangle}[1][]{\savebox{\@brx}{\(\m@th{#1\rangle}\)}%
  \mathclose{\copy\@brx\kern-0.5\wd\@brx\usebox{\@brx}}}
\makeatother
\newcommand{\lla}{\la\!\la}
\newcommand{\rra}{\ra\!\ra}

\begin{document}
 
\author{Fedor K.~Popov$^{1}$ and Yifan Wang$^{2}$
}
\affiliation{$^{1}$ Simons Center for Geometry and Physics, SUNY, Stony Brook, NY 11794, USA}
\affiliation{$^{2}$ Center for Cosmology and Particle Physics, New York University, New York, NY 10003, USA}

\title{ 
Factorizing Defects from Generalized Pinning Fields
}

\begin{abstract}

We introduce generalized pinning fields in conformal field theory that model a large class of critical impurities at large distance, enriching the familiar universality classes. We provide a rigorous definition of such defects as certain unbounded operators on the Hilbert space and prove that when inserted on codimension-one surfaces they factorize the spacetime into two halves. The factorization channels are further constrained by symmetries in the bulk. As a corollary, we solve such critical impurities in the 2d minimal models and establish the factorization phenomena previously observed for localized mass deformations in the 3d ${\rm O}(N)$ model.

\end{abstract}

\date{\today}

\pacs{}

\maketitle

\section{Introduction and Summary}

Extended operators defined on submanifolds of the spacetime constitute a fundamental component of modern quantum field theory (QFT), shaping our understanding of symmetry principles, phases of matter and strong coupling phenomena. 
When extended in time, they are also known as defects -- such as impurities and boundaries -- which exhibit rich dynamics, particularly in gapless bulk systems described by conformal field theories (CFTs). 
In cases where the CFT admits a gravity dual, defects correspond to strings, branes and nontrivial backreacted geometries, providing a valuable portal to investigate systematically such nonperturbative objects in quantum gravity. Hence it is essential to determine what defects and extended operators are admissible in CFT (and QFT more broadly) and to understand their corresponding dynamics.  

To this end, powerful monotonicity theorems have been established for defect renormalization group (RG) flows \cite{Affleck:1991tk,Friedan:2003yc,Jensen:2015swa,Casini:2018nym,Kobayashi:2018lil,Wang:2020xkc,Wang:2021mdq,Cuomo:2021rkm,Shachar:2022fqk,Casini:2023kyj}, ensuring that certain defects remain unscreened at large distances, while additional constraints from generalized symmetries are beginning to be explored \cite{Konechny:2019wff,Thorngren:2020yht,Aharony:2022ntz,Choi:2023xjw,Aharony:2023amq,Bhardwaj:2024igy,Choi:2024tri,Antinucci:2024izg}. Furthermore, conformal bootstrap techniques have been developed to directly probe the operator algebra data at the defect fixed point \cite{Billo:2016cpy}, while the study of defect fusion and the Casimir effect offers new perspectives on universal defect data via the framework of effective field theory \cite{Diatlyk:2024zkk,Diatlyk:2024qpr,Kravchuk:2024qoh,Cuomo:2024psk}. Nevertheless, explicit examples of nontrivial conformal defects remain scarce, and even with a well-defined short-distance description, their behavior at large distances remains unclear.

Here we introduce a large family of defects with the following simple UV definition in an arbitrary CFT in $d\geq 2$, 
\ie 
{\cal D}_h({\cO})\equiv \left[e^{h \hat \cO}\right]_{\rm ren}\,,\quad \hat \cO\equiv \int_{\Sigma_p} \cO(x) \,,
\label{genpindef}
\fe
where 
$\cO$ is a scalar operator of dimension $\Delta_\cO<p$ and $\Sigma_p$ is the defect worldvolume which we will often take to be an $\mR^p$ hyperplane in the Euclidean spacetime. Intuitively, a nonzero coupling constant $h$ in \eqref{genpindef} triggers a defect RG flow on $\Sigma_p$. With appropriate renormalization, this defines a $p$-dimensional flat defect in the CFT (similarly for $\Sigma_p=S^p$ which defines the spherical defect). Such a defect is expected to flow to a nontrivial conformal defect in the IR, which we denote as,
\ie 
{\cal D}({\cO})\equiv \lim_{h\to \infty} {\cal D}_h({\cO})\,.
\label{genpinconf}
\fe
This is because the monotonicity theorems requires the defect entropy (for $p$ odd) and conformal anomaly (for $p$ even) to decrease under defect RG, and they both vanish for the trivial defect.
These defects generalize the well-studied pinning field defects in the Ising and general ${\rm O}(N)$ CFTs where $\cO$ is taken to be the fundamental scalar field (measuring the local spin) and $h$ is the so-called background pinning field \cite{ParisenToldin:2016szc,Cuomo:2021kfm}.
We will thus refer to \eqref{genpindef} and \eqref{genpinconf} as the generalized pinning field defect and its conformal fixed point. In particular, our definition includes the defect RG flows studied by \cite{Krishnan:2023cff,Trepanier:2023tvb,Raviv-Moshe:2023yvq,Giombi:2023dqs,Diatlyk:2024ngd}  in the very same ${\rm O}(N)$ CFT but triggered by certain mass deformations on a surface. There it was observed based on large $N$ analysis and $\epsilon$-expansion that the generalized pinning field surface defect at $3d$ appears to factorize into conformal boundary conditions for two disconnected regions of the spacetime. Interestingly, this factorization also manifests in different ways depending on the sign of the pinning field  $h$, revealing the intricate phase structure
on  the surface, which includes a newly discovered extraordinary-log phase at strong surface coupling \cite{Metlitski:2020cqy,Krishnan:2023cff}. 

In this work, we focus on codimension-one ($i.e.$ $p=d-1$) generalized pinning field defects and establish this factorization property in great generality. In particular, this implies that there is no energy transmission across the interface in the IR.

Intuitively, this is because the conformal pinning defect \eqref{genpinconf} will take the form of an un-normalized projector on the CFT rigged Hilbert space $\cH$ (see around \eqref{eq:intro_rigged_Hilb}),
\ie 
\cD(\cO)=\bigoplus_\A|\A\ra \la \A|\,,
\label{intfact}
\fe
by projecting to the highest or lowest eigenvalues of the operator $\hat \cO$ depending on the sign of $h$, where the eigenbasis is denoted by $|\A\ra$.
The eigenstates $|\A\ra$ are scalars with respect to ${\rm SO}(d-1)$ and hence do not support a non-vanishing two-point function of the stress energy tensor $T_{\m\n}$ across the interface. Furthermore, by conformal symmetry, we expect these eigenstates to describe conformal boundary conditions of the CFT. 

 However one should be cautious with the above heuristic reasoning, since the conformal boundary conditions do not define normalizable states in $\cH$ and relatedly $\hat\cO$ may not be a healthy bounded operator on $\cH$. In the following we will formalize these points using results from the spectral theory of unbounded operators on Hilbert spaces, particularly the Gelfand triple construction. This framework allows us to precisely define generalized pinning field defects and prove their IR factorizations in Section~\ref{sec:fact}.

The factorization property enables us to determine the long-distance dynamics of such defects by leveraging the full spectrum of conformal boundary conditions, thereby rendering the problem significantly more tractable. For example, as we discuss in Section~\ref{sec:3deg}, this elucidates the conjectured factorization of the $3d$ ${\rm O}(N)$ CFT induced by the ${\rm O}(N)$-symmetric surface mass term in terms of known boundary conditions and also yields nontrivial predictions for more general surface mass deformations. 
As another example, the conformal boundary conditions of Virasoro minimal models in $2d$ were classified long time ago. However limited results are available for conformal (non-topological) line defects except for those in the Ising CFT. In Section~\ref{sec:2deg}, we analyze a large class of $2d$ pinning defects defined as in \eqref{genpindef}, demonstrating how factorization together with the monotonicity $g$-theorem and symmetry constraints allow us to completely nail down the IR conformal defect. 

In general, the symmetries of the CFT restrict the possible factorization channels in \eqref{intfact}. In particular, anomalies of the symmetries preserved by the defect will force degeneracies in the direct sum, as we discuss Section~\ref{sec:symconstraint}, and they play an important role in solving the pinning defects in $2d$ minimal models with examples in Section~\ref{sec:2deg} and Appendix~\ref{app:more2d}. 

As mentioned before, these generalized pinning defects are ubiquitous in CFT and thus it would be interesting to study their applications, as probes of strongly coupled systems. For instance, quantum chromodynamics (QCD) with a sufficiently large number of massless fermion flavors (within the conformal window) admit nontrivial Caswell-Banks-Zaks type fixed points \cite{Caswell:1974gg,Banks:1981nn}. Notably, the fermion mass operator  $\bar{\psi} \psi$ is conjectured to have a scaling dimension that interpolates between 2 and 3, varying from the strongly-coupled to the weakly-coupled ends of the conformal window \cite{Kaplan:2009kr,Gorbenko:2018ncu,Lee:2020ihn}. Our results predict the existence of nontrivial factorized interfaces in conformal QCD, arising from the corresponding pinning field flows, and may provide new ways for testing conjectures regarding the endpoint of the conformal window.
Furthermore, when viewed as an extended operator at fixed time, generalized pinning defects have been employed to model local quantum channels, capturing the effects of local decoherence and weak measurement on quantum critical states (see $e.g.$ \cite{Lee:2023fsk,Zou:2023rmw}). It would be interesting to explore the implications of our findings on the universality classes of such quantum channels. Finally, in CFTs with holographic duals, it would be interesting to investigate the dual description of codimension-one pinning defects, which may correspond to factorizing geometries involving new end-of-the-world branes \cite{Karch:2000ct,Karch:2000gx,DeWolfe:2001pq,Takayanagi:2011zk,Fujita:2011fp}.

The generalized pinning defects considered here are defined by defect RG flows from the trivial defect triggered by bulk local scalar operators. Immediate generalizations include such flows on more general topological defects, and triggered by more general operators, such as spinning local operators and even operators attached to nontrivial topological lines (see previous related works \cite{Bachas:2004sy,Runkel:2007wd,Bachas:2009mc,Kormos:2009sk,Manolopoulos:2009np,Gaiotto:2020fdr,Tavares:2024vtu} in $2d$). A further generalization amounts to considering RG flows on a slab, a fattened version of our setup here, and reduce to our setup when the intermediate phase in the slab is gapped in the IR. It will be interesting to understand general properties of such flows and the long distance behavior of the resulting defects.

\section{Factorization}
\label{sec:fact}

The main goal of this section is to prove the factorization property for \textit{generic} pinning flows,
\ie 
\cD\equiv \lim_{h\to \infty}\left[e^{h \hat{\mathcal{O}}}\right]_{\rm ren} = |\cB\ra\la \cB|\,,\label{eq:main_message}
\fe
where $|\cB\ra$ is a linear combination of Ishibashi states which are in one-to-one correspondence with scalar primary operators in the CFT \cite{Nakayama:2015mva}. This linear combination is further constrained by locality, and gives rise to a Cardy state in $2d$ \cite{Cardy:1989ir}. For this reason, we will also refer to $|\cB\ra$ as a Cardy state in general dimensions. As will soon become evident, the primary task lies in rigorously defining each component of \eqref{eq:main_message}. By \textit{generic}, we mean the flow does not preserve any extra symmetry and towards the end of the section we will discuss cases with extra symmetry, and how \eqref{eq:main_message} is modified accordingly.

For convenience, we will work in the Euclidean flat space with coordinates $\{\vec x,x_d\}$ such that the defect is planar and located at $x_d=0$. 
We first define the CFT Hilbert space $\mathcal{H}$ at $x_d=0$ in the N-S quantization \cite{Rychkov:2016iqz} as spanned by  
\ie
    |\phi_f\ra  = \sum_a \int\limits_{\mathbb{R}^d_+} d^d x\, f_a(x)\,\phi_a(x)|0\ra\,, \quad f_a: \mathbb{R}^d_+ \to \mathbb{C}\,,
    \label{genv}
\fe
where  $\phi_a(x)$ are normalized primary operators of dimension $\Delta_a$ with smearing factors $f_a$ on the half space $x_d>0$, and the index $a$ may also include spin indices. The
inner product with scalar operators $\phi_a$ in \eqref{genv} reads
\ie
    \braket{\phi_f|\phi_g} = \sum_a \int\limits_{\mathbb{R}^d_+} d^d x\, d^d y\,\frac{f_a^*(x) g_a(y)}{\left(|\vec x-\vec y|^2 + (x_d + y_d)^2\right)^{\Delta_a}}\,,
\fe 
and the generalization with spinning operators is obvious.
The smearing factors $f_a$ are constrained such that $|\phi_f\ra  \in \mathcal{H}$ has a finite norm, namely
$
||\phi_f|| \equiv  \sqrt{\braket{\phi_f|\phi_f}} < \infty$,
and the Hilbert space $\mathcal{H}$ is complete with respect to this norm.

Next we define the space to which the Ishibashi states belong, since they are not normalizable and therefore do not reside in the Hilbert space $\mathcal{H}$. We first consider a dense subspace $\Phi \subset \mathcal{H}$ ($i.e.$ its closure $\bar{\Phi} = \mathcal{H}$) spanned by finite linear combinations of primary operators and their descendants.
We then introduce the dual space $\Phi'$, consisting of all continuous linear functionals on $\Phi$ and equipped with the weak topology \cite{gel2013spaces, izrail1967generalized}. The full structure is captured by
a Gelfand triple (also known as a rigged Hilbert space) \cite{gel2013spaces, izrail1967generalized},
\ie 
\Phi \subset \mathcal{H} \subset \Phi'\,. \label{eq:intro_rigged_Hilb}
\fe  
Explicitly, an element $\ket{I} \in \Phi'$ is defined by the weak limit of elements in $\cH$, namely $|I\ra=\underset{n \to \infty}{\operatorname{w-lim}}\, |I_n\ra$ as below,
\ie 
 \exists \ket{I_n} \in \cH :   \forall \ket{\psi} \in \Phi\,,
   \lim_{n\to \infty} \braket{\psi|I_n} =  \braket{\psi|I}\,.
   \label{weaklimstate}
\fe 
In other words, even though $\ket{I}$ is not an element of the original Hilbert space, its overlap with any element in the dense subspace $\Phi$ is well-defined. For instance, if $\ket{I} = |\phi\rra$ is an Ishibashi state corresponding to a scalar primary operator $\phi$, 
\ie 
    \langle\psi|\phi(x) \rra = x_d^{-\Delta_\phi}\,.
\fe 
We emphasize that this overlap is not well-defined for general states in $\cH$ (see Appendix~\ref{app:ishibashioverlap}).
We have thus explicitly defined the RHS of equation~\eqref{eq:main_message}.

Now we define the regularization and  renormalization of the defect operator in \eqref{eq:main_message}. The first step is to regularize the operator $\hat \cO$ in the exponent so that it is self-adjoint and has well-defined spectral decomposition.
Here we consider a sequence of subspaces $\cH_\Delta \subset \cH $ which contain all conformal families of primary operators $\phi_\alpha$ with dimension $\Delta_\alpha \leq \Delta$. We assume that the CFT spectrum is discrete and does not contain accumulation points, thus $\cH_\Delta$ is a finite sum over conformal families and a Hilbert space itself. This regularization still contains an infinite number of states but explicitly preserves the conformal symmetry of the bulk theory which will be important in the subsequent analysis. 

To regularize $\hat \cO$ amounts to defining its action  on $\cH_\Delta$.  We define $\Phi_\Delta=\Phi\cap \cH_\Delta$ and introduce the following sesqulinear form on $\Phi_\Delta$ 
\ie 
(\phi_f,\phi_g)\equiv \braket{\mathcal{\phi}_f |\hat{\cO}| \mathcal{\phi}_g}\,,
\label{sesform}
\fe 
for $|\phi_f\ra,|\phi_g\ra \in \Phi_\Delta$ and the RHS is defined by the bulk OPE (see \eqref{app:eq:matrix_elements}).  Using  results in Appendix~\ref{app:existence_of_extension}, there is a unique self-adjoint operator $\hat{\cO}_\Delta$ with a dense domain in $\Phi_\Delta$ (thus also dense in $\cH_\Delta$) whose matrix elements coincide with that of $\hat \cO$ in \eqref{sesform}.
 The operator $\hat\cO_\Delta$ is self-adjoint but unbounded, as we describe below  
\footnote{Note that the matrix element as defined in \eqref{app:eq:matrix_elements} is IR finite only when $\Delta > \frac{d-1}{2}$
, a condition stemming from the use of the planar setup for the defect. However these divergences are naturally regularized in the spherical conformal frame, which is related to the planar frame by a conformal transformation. For the sake of conciseness, we choose to work in the planar frame with this understanding.}.

The conformal generators preserving the quantization surface are dilation $D$, translations $P_i$, special conformal transformations $K_i$ and ${\rm SO}(d-1)$ rotations $M_{ij}$. 
Acting on the Hilbert space $\cH$, we have explicitly
\begin{align}
     D& |\phi_f\ra  = \sum_a \int\limits_{\mathbb{R}^d_+} d^d x\left( x^\mu\partial_\mu + d - \Delta_a\right) f_a(x)\,\phi_a(x) |0\ra \,, \notag \\
     P_i& |\phi_f\ra  =  \sum_a \int\limits_{\mathbb{R}^d_+} d^d x\, \partial_i f_a(x)\,\phi_a(x)|0\ra\,.   \label{eq:sym}
\end{align}
Importantly, the unitary operator $U_b = e^{b D}$ implies,
\ie 
    &\braket{U_{b}\phi_f |U_{b}\phi_g} = \braket{\phi_f|\phi_g}\,,
    \\
    &\braket{U_{b}\phi_f|\hat{\mathcal{O}}_\Delta|U_{b}\phi_g} 
    = e^{b(d-1-\Delta_\cO)}\braket{\phi_f|\hat{\mathcal{O}}_\Delta|\phi_g}\,.
    \label{scaling}
\fe 
The operator $\hat{\mathcal{O}}_\Delta$ is thus clearly unbounded, since $U_{b}$ can be used to indefinitely increase its matrix elements between normalizable states
\footnote{One can show under the weak limit that 
	$\underset{b \to \infty}{\operatorname{w-lim}}\, U_b\ket{\psi} = 0$
	 for any $|\psi\ra \in \cH$ (see Section~\ref{sec:alternative_proof} for details).
}. 
By applying the Hellinger–Toeplitz theorem \cite{kolmogorov1957elements}, we deduce that the domain of $\hat\cO_\Delta$ is a proper subspace of the Hilbert space, $i.e.$ $\Phi_\Delta\subset \dom(\hat{\mathcal{O}}_\Delta) \subsetneq \mathcal{H}_\Delta$.
From \eqref{scaling} we also conclude $\hat{\mathcal{O}}_\Delta$ does not possess nontrivial eigenvectors in $\cH_\Delta$.

Nevertheless, \textit{generalized eigenvectors} of $\hat\cO_\Delta$ do exist in $\Phi'_\Delta$ (as in the Gelfand triple in \eqref{eq:intro_rigged_Hilb}) by invoking the Gelfand–Maurin theorem \cite{izrail1967generalized,gel2013spaces}. 
Specifically, we represent the self-adjoint operator $\hat{\mathcal{O}}_\Delta$ as
\ie 
    \hat{\mathcal{O}}_\Delta = \int\limits_{\mathbb{R}} \mu\, dE_{\hat{\mathcal{O}}_\Delta}(\mu), \, E_{\hat{\mathcal{O}}_\Delta}(\mu) E_{\hat{\mathcal{O}}_\Delta}(\lambda) = E_{\hat{\mathcal{O}}_\Delta}(\min(\mu,\lambda)) \notag \,.
\fe 
Here  $E_{\hat{\mathcal{O}}_\Delta}(\mu)$, known as projection measure, denotes the projection operator onto the closed subspace of $\cH_\Delta$ where the expectation value of  $\hat\cO_\Delta$ is less than or equal to
$\mu$.
Then the Gelfand-Maurin theorem  (see Chap 4 of \cite{izrail1967generalized}) assures the existence of the following derivative (almost everywhere) for any unit vector $\ket{e} \in \cH_\Delta$ with spectral density $\sigma_\lambda$ defined below,
\ie
{}&\ket{\lambda}  = \frac{d E_{\mathcal{O}_\Delta}(\lambda)}{d\sigma_\lambda}\ket{e} \in \Phi'_\Delta\,,~ d\sigma_\lambda \equiv \bra{e} dE_{\cO_\Delta} \ket{e}\,,
 \\ 
&{\rm s.t.}~
    \forall\,\ket{\psi}\in\Phi_\Delta\,,~
    \braket{\lambda|\psi}=\frac{d}{d\sigma_\lambda}\braket{e|E_{\hat{\mathcal{O}}_\Delta}(\lambda)|\psi}\,,  
\label{specdensity}
\fe
and $|\lambda\ra$ are the (generalized) eigenvectors of $\hat{\mathcal{O}}_\Delta$ in $\Phi'_\Delta$.

With subspaces $\mathcal{H}_\Delta(e) \equiv {\rm span}(\{ E_{\hat\cO_\Delta}(\lambda)\, e \})\subseteq \mathcal{H}_\Delta$, the Gelfand-Maurin theorem  states that if the Hilbert space decomposes as an orthogonal sum
$
\mathcal{H}_\Delta = \bigoplus_{\alpha \in \mathcal{A}}\mathcal{H}_\Delta(e_\alpha)
$
for a set of vectors $\{e_\alpha\}_{\alpha\in\mathcal{A}}$, we can define a spectral measure $d\sigma_{\lambda,\A}$ and the following decomposition holds
\begin{align}
    \hat{\mathcal{O}}_\Delta = \int\limits_{\mathbb{R}} \lambda\, d\sigma_{\lambda,\alpha} \ket{\alpha, \lambda, \Delta} \bra{\alpha, \lambda, \Delta}\,.
    \label{Ospec}
\end{align} 
Physically, $\A$ keeps track of extra symmetry charges that commute with $\hat \cO_\Delta$.

For the case of interest, $\hat{\mathcal{O}}_\Delta$ commutes with the momentum operators $P_i$ and ${\rm SO}(d-1)$ rotations, which do not mutually commute and we will choose to diagonalize $P_i$ below \footnote{If $\hat{\mathcal{O}}_\Delta$ also respects additional symmetries, these eigenstates are further diagonalized with respect to the corresponding charges.}. The generic condition stated below \eqref{eq:main_message} corresponds to 
assuming the existence of a cyclic vector $\ket{e}$ such that all linear combinations of the form
\ie 
E_{\mathcal{\cO}_\Delta} (\lambda) \prod\limits^{d-1}_{i=1} E_{P_i}(p_i) \ket{e}\,,
\fe 
span the Hilbert space $\cH_\Delta$, where $E_{P_i}$ denotes the projection measure for $P_i$, we can then decompose $\hat{\mathcal{O}}_\Delta$ in a basis of simultaneous eigenstates of $\hat{\mathcal{O}}_\Delta$ and $P_i$ as,
\begin{align}
   & P_i \ket{\lambda, \vec{p},\Delta} = p_i \ket{\lambda, \vec{p}, \Delta}, \quad 
    \hat{\mathcal{O}} \ket{\lambda, \vec{p}, \Delta} = \lambda |\lambda, \vec{p}, \Delta \ra\,.  \notag\\
    &\hat{\mathcal{O}}_\Delta = \int\limits_{\mathbb{R}} \lambda\, d\sigma_\lambda \int \frac{d^{d-1} \vec p}{(2\pi)^{d-1}}\ket{\vec{p}, \lambda} \bra{\vec{p}, \lambda}\,,  
\end{align}
This basis is complete, allowing us to express inner products of states in $\cH$ explicitly as below,
\ie
    \braket{\psi | \chi} = \int d\sigma_\lambda \int \frac{d^{d-1} \vec p}{(2\pi)^{d-1}} 
    \braket{\psi | \vec{p}, \lambda, \Delta} \braket{\vec{p}, \lambda, \Delta | \chi}\,.  
\fe 
For convenience, we introduce Wannier wave functions \cite{marzari2012maximally},
\ie 
    |\lambda,\vec{x} ,\Delta \ra \equiv \int \frac{d^{d-1} \vec p}{(2\pi)^{d-1}} e^{-i \vec{p} \cdot \vec{x}} |\lambda, \vec{p},\Delta \ra\,.
\fe
In this representation, the operators $D$ and $P_i$ act as
\begin{align}
    D |\lambda,\vec{x}, \Delta\ra =& \left(x^\mu \partial_\mu + (d-1-\Delta_{O}) \lambda \partial_\lambda\right)  |\lambda,\vec{x}, \Delta\ra\,, \notag\\
    P_i |\lambda,\vec{x}, \Delta\ra =&\, \partial_i |\lambda,\vec{x},\Delta\ra\,.
\end{align}
This allows us to determine the action of $K_i$ from the conformal algebra,
\ie 
  &  K_j |\lambda, \vec{x}, \Delta\ra   = \left(2 (d-1-\Delta_\cO) x_i \lambda\partial_\lambda \right.\\
  &\left.+ (2 x_i (x\cdot\partial) - x^2 \partial_i)\right)|\lambda, \vec{x}, \Delta\ra\,.  
\fe 
Now we consider the dense subset $\mathcal{H}_{\mathcal{D}_\Delta} \subset \mathcal{H}_\Delta$, where for any $\ket{f} \in \mathcal{H}_{\mathcal{D}_\Delta}$ the following limit exists
\begin{align} 
    \exists \lim_{\lambda\to\infty} \braket{\lambda, \vec{x}, \Delta|f } \Rightarrow \ket{\vec{x}, \Delta} = \underset{\lambda \to \infty}{\operatorname{w-lim}}\, |\lambda,\vec{x}, \Delta\ra\,. 
    \label{wlx}
\end{align}
In general this weak limit might not exist or could simply be zero \footnote{An example where this limit does not exist is discussed in Appendix~\eqref{app:free_field}.}. Physically, this amounts to demanding a well-defined IR limit of correlation functions of $\hat \cO$ (and the defect) with bulk local operators. 
It is straightforward to check that the limit states $\ket{\vec{x},\Delta}$ form a representation of the algebra $\{P_i, D, K_j,M_{ij}\}$. For instance, for any state  $\ket{f} \in \mathcal{H}_{\mathcal{D}_\Delta}$
\ie
 & 
  \lim_{\lambda \to \infty }\braket{f|e^{b D}| \lambda, \vec{x} }
 = \lim_{\lambda \to \infty }\braket{f| \lambda e^{b(d-1-\Delta)}, e^{b} \vec{x}}
   \\
   &\quad =\braket{f|e^{b} \vec{x}}  \quad 
 \Rightarrow \quad e^{bD} \ket{\vec{x}} = \ket{e^{b}\vec{x}}\,.
\fe 
where $\Delta$ dependence is  implicit.
From Appendix~\ref{sec:important_lemma}, we show that for all $\vec{x}$, the limit is the same in \eqref{wlx},  
\ie 
\ket{\vec{x}, \Delta} = \ket{\mathcal{B}, \Delta}\,,
\label{xtoishi}
\fe 
 which is a linear combination of Ishibashi states corresponding to scalar primaries.

After having analyzed the spectrum of $\hat \cO_\Delta$, we are ready to deal with its exponential in \eqref{eq:main_message},
\ie 
    \mathcal{D} \equiv     \underset{\Delta\to\infty}{\operatorname{w-lim}}\, \underset{h\to\infty}{\operatorname{w-lim}}  \left[ e^{h \hat{\mathcal{O}}_\Delta} \right]_{\rm ren} \,,
\fe 
by introducing a cutoff $\Lambda$ on the spectrum  of $\hat{\cO}_\Delta$ and then taking weak limits. Before these limits, the renormalized defect is 
defined via minimal subtraction as
\begin{align} 
    \mathcal{D}_{\Delta}(h,\Lambda) \equiv  &\label{eq:reg_procedure} \\
    \cN\int\limits^{\Lambda}_{-\infty} d\sigma_\lambda \frac{e^{h (\lambda - \Lambda) }}{h} \int \frac{d^{d-1}\vec p}{(2\pi)^{d-1}} &\ket{\lambda, \vec{p}, \Delta} \bra{\lambda, \vec{p}, \Delta}. \notag
\end{align}
with a normalization factor $\cN$ which we will fix later. 
The following weak limit clearly exists,
\ie 
   \underset{h\to\infty}{\operatorname{w-lim}}\,   \mathcal{D}_{\Delta}(h,\Lambda) = \cN\left.\frac{d\sigma_\lambda}{d\lambda} \right|_{\lambda=\Lambda} \\
   \times \int \frac{d^{d-1} \vec p}{(2\pi)^{d-1}} \ket{\Lambda, \vec{p}, \Delta} \bra{\Lambda, \vec{p}, \Delta}\,,
   \label{weakhlim}
\fe 
where we have used that
\ie 
    \lim_{h\to\infty} h e^{h(\lambda-\Lambda)} \theta(\Lambda-\lambda) = \delta(\Lambda - \lambda).
\fe 
Taking weak limit $\Lambda\to \infty$, using \eqref{wlx} together with \eqref{xtoishi} and then $\Delta\to \infty$, we obtain
\ie 
\cD=|\cB\ra  \la \cB|\,,\quad \ket{\mathcal{B}} \propto \underset{\Delta \to \infty}{\operatorname{w-lim}} \ket{\mathcal{B}, \Delta} \,,
\label{weakDD}
\fe
where we have absorbed the normalization constants in \eqref{weakhlim} and the IR volume factor into $|\cB\ra$, which is a linear combination of Ishibashi states.  By locality, we expect $|\cB\ra$ to be an indecomposable Cardy state of the CFT and its normalization is fixed this way.

Let us comment on the extensions to \eqref{eq:main_message} when the pinning flow preserves additional global symmetries. This happens when $\hat \cO$ commutes with the corresponding topological defects.  Correspondingly, the spectrum of $\hat \cO$ 
has degeneracies due to the symmetry action (fusion with the topological defects). When the symmetry preserved is finite, this leads to a finite direct sum of factorized interfaces in the IR for the pinning defect,
\begin{align} 
    \mathcal{D} = \sum_\alpha |{\mathcal{B}_\alpha}\ra \la \mathcal{B}_\alpha|\,, \label{eq:decompose}
\end{align} 
where the individual Cardy states $|\cB_\A\ra$ furnish a representation of the symmetry. We will see concrete examples in Section~\ref{sec:examples}.

If the symmetry is continuous, we have instead an integral over the factorization channels,
\ie 
  \mathcal{D} = \int d\A |{\mathcal{B}_\alpha}\ra \la \mathcal{B}_\alpha|\,, \label{eq:decompose2}
\fe
for a spectral measure $d\A$ invariant under the symmetry. For example if $\A$ labels the elements of a continuous group symmetry $G$, $d\A$ is proportional to the Haar measure on $G$. In the case of spontaneous $G$-symmetry breaking on the interface, this measure arises from the path integral over the Goldstone bosons localized at the interface. For $3d$, there is no spontaneous continuous symmetry breaking  \cite{Cuomo:2023qvp}, but this form of factorization \eqref{eq:decompose2} persists (see Section~\ref{sec:3deg}).

The factorized form of the pinning defect derived here makes their IR fusion product $\circ$ defined in \cite{Diatlyk:2024zkk} (see also \cite{Bachas:2007td,Bachas:2013ora,Soderberg:2021kne,SoderbergRousu:2023zyj,Diatlyk:2024qpr,Kravchuk:2024qoh,Cuomo:2024psk}) more transparent. In particular, the defect self-fusion follows an idempotent rule,
\ie 
\cD \circ \cD (\Sigma) = \cS(\Sigma)\cD(\Sigma)\,,
    \label{projfusion}
\fe 
where $\Sigma$ is the defect worldvolume and $\cS$ denotes the fusion coefficient theory (CFT) on $\Sigma$ and $\cS(\Sigma)$ is its partition function. Here we have used that the boundary states $|\cB_\A\ra$ in the factorization of $\cD$ are related by symmetries and the Casimir energy densities $\cE_{\A\B}$ for a pair of such boundaries defined by
\ie 
 \la \cB_\A| e^{-z P_d}  |\cB_\B\ra \xrightarrow{z\to 0} e^{-{A\over z^{d-1}}\cE_{\A\B}}\,,
 \fe 
 with IR regulator $A$ for the defect volume, satisfies
 \ie 
\cE_{\A\B}=\cE_{\B\A} \geq \cE_{\A\A}=\cE_{\B\B}\,,
\fe
which can be derived using reflection positivity \cite{Diatlyk:2024qpr}.

Let us note that our approach shares similarities with Hamiltonian truncation ($e.g.$ truncated conformal space approach or TCSA) \cite{Yurov:1989yu}. However, instead of diagonalizing the full Hamiltonian $ H_\lambda \equiv K_d + P_d + \lambda \hat{\cO}$, we diagonalize only the perturbation $\hat{\cO}$. This distinction arises from the fact that we are primarily interested in the instantaneous action of the perturbation rather than the deformation of the whole spectrum in such theories. Nevertheless, if the perturbation induces a gap in the bulk, the eigenstates of  $\hat{\cO}$ can be used as a variational ansatz to approximate the ground state of the interacting theory \cite{Cardy:2017ufe,Lencses:2018paa, Konechny:2023xvo}, effectively neglecting the $K_d + P_d$ term in the full Hamiltonian $H_\lambda$. In contrast,  to produce a gapless theory in TCSA, the original Hamiltonian $K_d + P_d$ is important to balance the perturbation $\lambda \hat{\mathcal{O}}$, and the connection to our pinning flow becomes more obscure.

\section{Symmetry Constraints}
\label{sec:symconstraint}

The factorization channels that describe the pinning defect in the IR are further constrained by generalized global symmetries in the bulk CFT. In modern understanding, while local operators transform as ordinary charges of standard 0-form symmetry, extended defects transform in higher representations of the generalized symmetries (see $e.g.$ \cite{Delmastro:2022pfo,Brennan:2022tyl,Bartsch:2022mpm,Bartsch:2022ytj,Bhardwaj:2023wzd,Bhardwaj:2023ayw,Copetti:2024onh}). Intuitively, such a representation contains information about the fusion of the topological symmetry defects with the (non-topological) defects and the topological junctions between them, subject to consistencies of isotopy invariance. To make the following discussion concrete and self-contained, we will focus on $2d$, in which case the CFT symmetry is described in part by a fusion (sub)category $\cC$ and the symmetry properties of general line defects are captured by a bimodule category $\cM$ over $\cC$. 
Physically, the simple objects label indecomposable defects, and include their cousins from fusion with the topological defects $\cL_i\in \cC$. As will explain below, the symmetry properties of pinning defects are described by special bimodule categories. The generalization to higher dimensions is conceptually similar though the details are incomplete.

We define a \textit{$\cC$-symmetric defect} $\cD$ to be such that it is transparent to, and thus commutes with, all topological defects $\cL_i\in \cC$,
\ie
\cL_i  \circ \cD =\cD \circ \cL_i\,.
\label{comm}
\fe
The trivial defect is obviously $\cC$-symmetric, and so are the generalized pinning defects \eqref{genpindef} as long as the defining local operator $\cO$ commutes with $\cL_i$. As we will see, the $\cC$-symmetric condition produces nontrivial constraints on the possible IR behaviors of the defects, analogous to what was found for symmetric RG flows in the bulk CFT
\cite{Chang:2018iay}, such as \textit{symmetry enforced degeneracies} in the IR factorization. 
\begin{figure}
    \centering
  \begin{tikzpicture}[scale = 0.5]
    \definecolor{lightblue}{RGB}{173, 216, 230}
    \definecolor{lightorange}{RGB}{255, 204, 153}
    
    \fill[lightblue] (6+.2,2) rectangle (8,-2);
    \fill[lightblue] (12,2) rectangle (14-.2,-2);
    \fill[lightorange,opacity=0.4] (8,-2) rectangle (12,2);
    \node at (7+.1, 0) { \Large$\mathcal{T}$};
    \node at (13-.1, 0) {\Large $\mathcal{T}$};
    \node at (10, 0) {\small $ h' \int d^{d} x \cO(x)$};
    \draw[<-] (4.1, 0) -- (5.9, 0)  node[midway,above]{\small strip RG};
    \fill[lightblue] (-4+.2,2) rectangle (-2,-2);
    \fill[lightblue] (2,2) rectangle (4-.2,-2);
    \fill[lightorange,opacity=0.8] (-2,2) rectangle (-0.75,-2);
    \fill[lightorange,opacity=0.8] (2,2) rectangle (0.75,-2);

    \node at (-3+.1, 0) {\Large $\mathcal{T}$};
    \node at (3-.1, 0) {\Large $\mathcal{T}$};
    
    \draw[thick, orange] (-0.75,2) -- (-0.75,-2);
    \draw[thick, orange] (0.75,2) -- (0.75,-2);
    
    \node at (0, 0) {\small $\mathcal{T}_{\rm gap}$};
    \node[orange] at (0, -2.5) {\small $|\nu\rangle\langle \nu|$};
    \draw[->] (0,-3.2) -- (0,-4.5)  node[midway,right]{\small topological fusion};

    \draw[<-] (10,-3) -- (10,-4.5)  node[midway,right]{\small resolve} ; 
    \fill[lightblue] (-3,-5) rectangle (-0.5,-9);
    \fill[lightblue] (3,-5) rectangle (0.5,-9);
    \node at (-2, -7) {\Large $\mathcal{T}$};
    \node at (2, -7) {\Large $\mathcal{T}$};
    \draw[thick, blue] (-0.5,-5) -- (-0.5,-9);
    \draw[thick, blue] (0.5,-5) -- (0.5,-9);
    \node[blue] at (0,-10) {\small $|\mathcal{B}_\nu\rangle \langle \mathcal{B}_\nu|$};
    \node at (0.75,-10) {\small $$};
    \draw[<-] (3.5,-7) -- (7.5,-7) node[midway, above] {\small $\mathcal{D} = \underset{\nu}{\bigoplus} \ket{\mathcal{B}_\nu} \bra{\mathcal{B}_\nu}$};
    \node at (5.5,-7.7) {\small defect RG};

    \fill[lightblue] (8,-5) rectangle (10,-9);
    \fill[lightblue] (12,-5) rectangle (10,-9);
    \draw[thick] (10,-5) -- (10,-9);
    \node at (9, -7) {\Large $\mathcal{T}$};
    \node at (11, -7) {\Large $\mathcal{T}$};
     \node at (10, -10) {$\mathcal{D} = e^{h \int d^{d-1} x \cO(x)}$};

\end{tikzpicture}
\caption{
IR factorization of the pinning flow via the strip flow by resolution. Here it is assumed that strip flow of the CFT $\cT$ ends in a gapped phase (orange region) described by TQFT $\cT_{\rm gap}$ with Cardy branes $|\n\ra$. }
\label{fig:reg_def}
\end{figure}
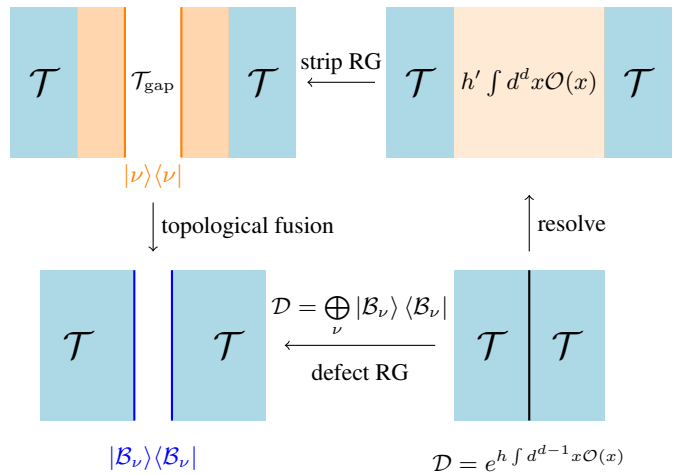

Let us offer some intuition behind this connection between defect and bulk symmetric RG flows as in Figure~\ref{fig:reg_def}, which also gives another way to see the factorization phenomena we have proven in Section~\ref{sec:fact}. The pinning flow defined in the UV by \eqref{genpindef} can be alternatively described by turning on the same operator $\cO$ on a strip of thickness $\D$ in the limit of $\D/L \ll 1$ where $L$ is the typical size for a bulk observable (possibly with fine-tuning when the OPE of $\cO$ generate more relevant operators).  If we follow the bulk RG flow on the strip (namely taking $\m \D\gg 1$ with $\m$ as the scale of the flow) and suppose this flow gaps the bulk CFT, we conclude the strip region is effectively described by a TQFT $\cT_{\rm gap}$. A TQFT can be cut-open by inserting a complete basis of Cardy branes $|\n\ra$ and further fusing each of these topological boundaries onto the interface between $\cT$ and $\cT_{\rm gap}$ produce a simple boundary condition $|B_\n\ra$ for the CFT $\cT$. Assuming that the orders of the limits commute, we conclude that the pinning defect in the IR is described by the factorized product
\ie 
\cD(\cO)= \underset{\n}{\bigoplus} |\cB_\n\ra \la \cB_\n|\,,
\label{symfact}
\fe
where $\n$ labels the Cardy branes in the TQFT $\cT_{\rm gap}$. The Cardy branes are in one-to-one correspondence with the vacua of the TQFT (that obey cluster decomposition) \cite{Komargodski:2020mxz,Huang:2021zvu}, and these vacuum degeneracies are consequences of generalized anomalies for $\cC$. Here we see how such constraints translate to the pinning defects straightforwardly (in terms of the factorization channels).

Note that in \eqref{symfact} we have not specified the conformal boundaries $|B_\n\ra$, which depend on the interface between $\cT$ and $\cT_{\rm gap}$. These are known as RG interfaces and have been determined explicitly in the $2d$ Ising and tricritical Ising CFTs \cite{Konechny:2016eek,Cardy:2017ufe,Lencses:2018paa,Konechny:2023xvo}.
We will see how to bootstrap this information efficiently in Section~\ref{sec:2deg} with general criteria from monotonicity theorems and the $\cC$-symmetric condition.

As a sanity check, the indecomposable $\cC$-bimodule category $\cM$ labeling the conformal pinning defect should be orientation-reversal invariant 
\ie 
\cM \cong \overline{\cM}\,,
\fe
and compatible with the idempotent fusion rule \eqref{projfusion}, naturally in terms of the relative Deligne tensor product over $\cC$ \cite{etingof2010fusion,Douglas_2019},
\ie 
\cM\boxtimes_\cC \cM \cong n\cM\,,
\label{topfusionbim}
\fe 
where $n\in \mZ_+$ is the fusion multiplicity. The physical meaning of $n$ is better understood in the case of topological fusion as capturing an emergent 1d TQFT, but it is somewhat mysterious in the non-topological fusion considered here.  
 One way to connect the two (and give a physical definition of $n$ in the present case) is 
by considering a bulk $\cC$-symmetric RG flow to a gapped phase. Note that the symmetry category $\cC$ itself has a canonical $\cC$-bimodule category structure and is known as the canonical (trivial) $\cC$-bimodule category because it's the identity with respect to the relative tensor product. In particular it trivially solves \eqref{topfusionbim} with $n=1$ as expected because it labels the trivial (transparent) defect.

The indecomposable $\cC$-symmetric TQFTs, describing $\cT_{\rm gap}$ above, are in one-to-one correspondence with indecomposable $\cC$-module categories $\cM_{\rm gap}$  \cite{Thorngren:2019iar,Huang:2021zvu}. The simple objects of $\cM_{\rm gap}$ are precisely the Cardy branes $|\n\ra$. 
In fact, $\cM_{\rm gap}$ is a $(\cC,\cC_{\rm dual})$-bimodule category where $\cC_{\rm dual}$ is the dual symmetry under a generalized gauging in $\cC$ \cite{Thorngren:2019iar,Huang:2021zvu}. Naively, one may expect that the defect $\cC$-bimodule category $\cM$ is given by the relative tensor product of ${\cM}_{\rm gap}$ and its orientation reversal $\overline{\cM}_{\rm gap}$ over $\cC_{\rm dual}$,
\ie 
\cM_{\rm gap}\boxtimes_{\cC_{\rm dual}}  \overline{\cM}_{\rm gap}\cong \cC\,,
\fe
which gives the trivial bimodule category $\cC$, shared by the trivial defect as mentioned above \cite{etingof2010fusion}. However in general, $\cM$ is given by a quotient of $\cC$, realized by a surjective $\cC$-bimodule functor $\cC \to \cM$, reflecting the fact that some topological defects in $\cC$ may be absorbed by the defect, 
\ie 
\cL_i \circ \cD =\cD \circ \cL_i=\cD\,.
\label{symabs}
\fe 
 
Finally, so far we have focused on the symmetries that commute with the defect as in \eqref{comm}. There are further constraints from the broken symmetries (knowing exactly how the symmetry is broken). For example, when the symmetry defect $\cL$ anti-commutes with $\cO$, one can infer the following fusion rule between the pinning defects (with opposite signs of $h$) and $\cL$,
\ie 
\cL \circ  \cD_\pm(\cO)  = \cD_\mp(\cO)\circ \cL\,.
\label{anticom}
\fe
which further implies that the IR $g$-function is independent of the sign of the pinning flow \cite{Konechny:2019wff}. In the special case of invertible $\cal L$, namely its orientation reversal $\overline{\cL}$ defines the fusion inverse,
\ie 
\cL \circ \overline{\cL}=\id\,,
\label{anticomfus}
\fe
one conclude that
\ie 
\cD_\pm (\cO)=\cL \circ \cD_\mp (\cO) \circ \cL\,.
\label{invertibleconjugation}
\fe

\section{Examples}
\label{sec:examples}

In this section, we combine the factorization property and symmetry constraints to solve the IR behavior of generalized pinning defects \eqref{genpindef} in $2d$ and $3d$ CFTs. 

\subsection{Pinning Flows in 2d Virasoro minimal model CFT}
\label{sec:2deg} 

While boundary conditions of Virasoro minimal model CFTs are well-known and in one-to-one correspondence with scalar primaries, much less is known about conformal defects. In general that would involve analyzing boundary conditions for the doubled theory $\cT\times \overline{\cT}$ from folding the minimal model $\cT$ at the defect line. The resulting theory has conformal central charge $c>1$ and the boundary conditions are not classified with the exception for the Ising CFT \cite{Oshikawa:1996dj}. Nonetheless, with the general results presented here, we will have definite predictions for the IR behaviors of generalized pinning defects in $\cT$.

As a warm up, let us consider the Ising CFT, in which case the only operator of dimension $\Delta<1$, which could be a candidate for defining a pinning flow, is the spin operator $\sigma$ of dimension $\Delta={1\over 8}$. This operator breaks all the symmetries in the Ising CFT and produces a trivially gapped phase when turned on in the bulk. Consequently, we conclude that the IR defect takes the form of \eqref{symfact} with a single factorization channel $\cD(\sigma)=|\cB\ra\la\cB|$, into a fixed boundary condition $|\cB\ra$ for the Ising CFT. Furthermore $g$-theorem requires the corresponding boundary to have $g$-function $g_\cB<1$, for which the only possibilities are the Dirichlet boundaries $|\pm \ra$ with $g_{\pm}={1\over \sqrt{2}}$.
This fixes uniquely the conformal pinning defect in the Ising CFT to be 
\ie 
\cD_\pm (\sigma)= 
    |\pm \ra \la \pm| \,,
\label{isingflows}
\fe
depending on the sign of the pinning field $h$. As a consistency check, since the operator $\sigma$ is odd under the $\mZ_2$ global symmetry, the two flows in \eqref{isingflows} must be related by conjugation \eqref{invertibleconjugation} with the symmetry defect $\eta$,
\ie 
D_+(\sigma)=\eta \circ D_-(\sigma) \circ \eta\,,
\fe 
which is indeed satisfied since $\eta |\pm\ra =|\mp \ra$ under defect-boundary fusion.  

As a more interesting application, we now consider the tricritical Ising CFT which contains three operators of dimension $\Delta<1$: $\epsilon_{{1\over 10},{1\over 10}}\,,
\sigma_{{3\over 80},{3\over 80}}
$ and $\sigma'_{{7\over 16},{7\over 16}}$. The global symmetries of the theory is described by,
\ie 
\cC={\rm Ising}\boxtimes {\rm Fib}\,,~{\rm Irr}({\rm   Ising})=\{\id, \eta,\cN\}\,,~{\rm Irr}({\rm   Fib})=\{\id, W\}\,,
\fe
where $\cN$ is the duality defect with quantum dimension $\la \cN\ra=\sqrt{2}$ and $W$ is the nontrivial Fibonacci symmetry defect with $\la W\ra={\sqrt{5}+1\over 2}$ and they satisfy the topological fusion rules,
\ie  
\cN\circ \cN=\id\oplus \eta\,,~\cN\circ \eta=\eta\circ \cN=\cN\,,~W\circ W=\id\oplus  W\,.
\fe
The tricritical Ising CFT has 6 indecomposable boundary conditions and they are related by fusion with the topological defects as follows
\ie 
&|+\ra\,,|-\ra=\eta|+\ra\,,|0\ra=\cN|+\ra\,,
\\
&|d\ra=W|0\ra\,,|0-\ra=W|+\ra\,,|+0\ra=\eta|0-\ra\,.
\label{timbcs}
\fe
In particular, the most stable boundary conditions $|\pm\ra$ has $g$-function $g_{|\pm\ra}=\sqrt{\sin {\pi\over 5}\over 5}$
and the $g$-functions for 
other boundaries follows from fusion and the quantum dimensions of the topological defects.

It is well-known that the bulk flows with respect to the three operators $\cO=\epsilon,\sigma,\sigma'$ are massive \cite{Lassig:1990xy}. We summarize the symmetries $\cC$ preserved by such flows and the corresponding IR gapped phases in Table~\ref{tab:TIMflows}. The corresponding conformal pinning defects are then fixed in terms of conformal boundaries \eqref{timbcs} by the factorization channels in \eqref{symfact}, together with the $g$-theorem \footnote{For the pinning flow with $\cO=\sigma$, to completely fix the answer, we have additionally assumed that the unique factorization channel is via a stable boundary state. This also agrees with the findings of \cite{Cardy:2017ufe,Lencses:2018paa,Konechny:2023xvo}.}. The results are listed in Table~\ref{tab:TIMflows} where we also include the bimodule categories $\cM$ labeling the conformal defects in the IR. There Vec denotes symmetry-absorbing cases \eqref{symabs} and all other cases coincide with the canonical (trivial) bimodule category $\cC$. The boundary states that participate in the factorization channels also agree with the RG interfaces identified in 
 \cite{Cardy:2017ufe,Lencses:2018paa,Konechny:2023xvo}.

 \onecolumngrid
 \begin{center}
\begin{table}[!htb]
    \centering
    \begin{tabular}{|c|c|c|c|c|c|c|c|}
                \hline
                \multirow{2}*{~Operator \textbf{$\cO$} }&  \multirow{2}*{{Symmetry $\cC$}} & \multicolumn{2}{
                c|}{Bulk flow} &  
               \multicolumn{2}{
                 c| }{Defect flow}& \multicolumn{2}{
                 c| }{Bimodule category $\cM$}
                \\ \cline{3-8}
              & & $h>0$ & $h<0$ & $h>0$ & $h<0$  & $~~~~h>0~~~~$ & $h<0$\\
                \hline\hline
                $\epsilon$ &  ${\rm Vec}_{\mZ_2}$ & SSB$_2$ & Trivial & $|+\ra\la +|\oplus |-\ra\la-|$ & $|0\ra\la 0|$  & ${\rm Vec}_{\mZ_2}$ & ${\rm Vec}$ \\
                \hline
                $\sigma$ &   ${\rm Vec}$ & Trivial & Trivial  & $|+\ra\la +|$ & $|-\ra\la -|$  & ${\rm Vec}$  & ${\rm Vec}$\\
                \hline
                $\sigma'$ & Fib & SSB$_2$   & SSB$_2$ & $|+\ra\la +|\oplus |0-\ra\la-0|$ & $|-\ra\la -|\oplus |{+}0\ra\la +0|$ 
                 & Fib  & Fib\\
                \hline
            \end{tabular}
    \caption{The phase structure in the bulk and on the pinning defect depending on the deforming operator $\cO$. We use SSB$_k$ to denote a symmetry breaking phase with $k$ degenerate ground states.}
    \label{tab:TIMflows}
\end{table}
\end{center}
 \twocolumngrid

It is straightforward to extend this analysis to other minimal models. In Appendix~\ref{app:more2d}, we give more examples. There we also explain that certain generalizations of the pinning flows considered here, with a nontrivial topological defect in the UV deformed by twisted operators, can be solved in a similar way.

\subsection{Pinning Flows in 3d ${\rm O}(N)$ CFT}
\label{sec:3deg}

In the 3d ${\rm O}(N)$ CFT, there are two simple kinds of surface pinning flows, by considering either the fundamental scalar operator $\phi_I$ or the mass operators (either an ${\rm O}(N)$ scalar or tensor), which all have dimension $\Delta<2$ (see Table~\ref{tab:ONopdim} for a summary).

\begin{table}[!htb]
    \centering
    \setlength\extrarowheight{2pt}
    \begin{tabular}{|c|c|c|@{}c@{}|}
    \hline   $N$  & $\Delta_\phi$ & $\Delta_S$ &  $\Delta_T$ \\
\hline 
1 & 0.5181489(10) & 1.412625(10) &  \cellcolor{gray!25} \\
2 & 0.519088(22) & 1.51136(22)  & ~1.23629(11)~~\\ 
Large $N$ & ${1\over 2}+{4\over 3\pi^2}{1\over N}+\dots$ & 
         $2-{32\over \pi^2}{1\over N}+\dots$ & ~$1+{32\over 3\pi^2}{1\over N}+\dots$\\[0.5ex] 
\hline
    \end{tabular}
    \caption{The scaling dimensions of the fundamental scalar, the singlet and tensor mass operators in the ${\rm O}(N)$ CFT.  See \cite{Henriksson:2022rnm} and references therein.}
    \label{tab:ONopdim}
\end{table}

From our general result in Section~\ref{sec:fact}, the corresponding IR pinning defect will factorize into conformal boundary conditions of the ${\rm O}(N)$ CFT. In particular, the ${\rm O}(N)$ symmetric boundary conditions are classified into the ordinary, the special and the extraordinary-log classes \cite{Diehl:1996kd,Metlitski:2020cqy}. The first two cases are more familiar and can be thought of as interacting versions of the Dirichlet and Neumann conformal boundary conditions for free theory. The last case is more exotic, not strictly speaking conformal, due to logarithmic behaviors in correlation functions, and turns out to be closely related to the normal boundary condition $|\vec n\ra$ labeled by $\vec n \in {\rm S}^{n-1}$ which has nonzero one-point function with the fundamental scalar $\vec n\cdot \phi$ and breaks the ${\rm O}(N)$ symmetry \cite{Metlitski:2020cqy}. This extraordinary-log boundary class exists via a marginally irrelevant coupling between $|\vec n\ra$ and $N-1$ 2d Goldstones $\vec \pi$ for $2\leq N<N_{\rm cr}\approx 5$ \cite{Metlitski:2020cqy,Toldin:2021kun,Padayasi:2021sik} and have a generalization for interface with no upper critical $N$ \cite{Krishnan:2023cff}.
 
Let us first consider the case of ${\rm O}(N)$ symmetric pinning flows, namely $\cO=(\phi^2)_S$. Based on large $N$ analysis, it was proposed in \cite{Krishnan:2023cff} that for $h<0$, the pinning defect factorizes into the ordinary boundary conditions,
\ie 
D_+((\phi^2)_S)=|{\rm Ord}\ra\la {\rm Ord}|\,.
\fe
This is also supported by resumming perturbative results for scaling dimensions on the surface defect in $d{=}4{-}\epsilon$ dimensions \cite{Diatlyk:2024ngd} and comparing to numerical results for $|{\rm Ord}\ra$ \cite{Toldin:2023fny,Zhou:2024dbt}.
This factorization is in agreement with our general result since the positive mass ($h<0$) deformation gaps the theory to a single ground state (thus one factorization channel) when turned on in the bulk. The case with $h>0$ is more interesting, as the negative bulk mass deformation leads to a gapless phase with $N{-}1$ Goldstone modes, and correspondingly, the proposal of \cite{Krishnan:2023cff} suggests that the defect is described by factorized normal boundary conditions $|\vec n\ra\la \vec n|$ weakly coupled to $N{-}1$ Goldstone modes. Effectively, defined as an operator on the CFT Hilbert space, in the IR limit, we have
\ie 
D_-((\phi^2)_S)\propto \int_{{\rm S}^{n-1}} d\vec n |\vec n\ra\la \vec n|\,,
\fe 
which again agrees with our general results. 

For ${\rm O}(N)$ breaking pinning flows, such as that by the fundamental scalar, the natural expectation is a factorization into normal boundary conditions, 
\ie 
D_\pm (\vec n \cdot \phi) =|{\pm}\vec n \ra \la \pm \vec n|\,.
\fe
preserving the residual ${\rm O}(N{-}1)$ symmetry.
For ${\rm O}(N)$ breaking surface mass deformations, the factorization property, together with previous work for such surface defects in $d{=}4{-}\epsilon$ dimensions \cite{Trepanier:2023tvb,Raviv-Moshe:2023yvq} suggest new boundary universality classes in the ${\rm O}(N)$ CFT. It would be interesting to study them from complementary methods such as the fuzzy sphere approach \cite{Dedushenko:2024nwi,Zhou:2024dbt}.

 \section*{Acknowledgements}
We thank  Gabriel Cuomo, Zohar Komargodski, Justin Kulp,  Brandon Rayhaun, Max Metlitski, Yu Nakayama, Ananda Roy, Ingo Runkel and Slava Rychkov for helpful questions and discussions. 
YW thanks
Kyushu University Institute for Advanced Study and RIKEN Interdisciplinary Theoretical and Mathematical Sciences Program for hospitality and discussions at the ``Kyushu IAS-iTHEMS workshop: Non-perturbative methods in QFT" during the completion of this work.
The work of YW was
supported in part by the NSF grant PHY-2210420 and by the Simons Junior Faculty Fellows program.

\bibliographystyle{apsrev4-1}
\bibliography{defect}

\newpage
\onecolumngrid

\appendix

\section{More pinning flows and generalizations in $2d$ CFT}
\label{app:more2d}

In this appendix, to demonstrate the consequences of our general results on factorization and symmetry properties for pinning flows, 
we provide more examples in $2d$ CFT. We will also discuss generalizations that involve flows from nontrivial topological defects by turning on operators in twisted sectors. 

\subsection{Pinning flows in diagonal Virasoro minimal models}

We will consider three types of pinning field flows in the diagonal Virasoro minimal model $M_{m,m+1}$ with $m\geq 3$ which describes the multicritical point of the Landau-Ginzburg model with $\varphi^{2(m-1)}$ potential. They are defined by taking the operator $\cO$ in \eqref{genpindef} to be one of the following three scalar primaries,
\ie 
\phi_{2,2}\sim \varphi \,,~\phi_{1,2}\sim :\varphi^{m-2}:\,,~\phi_{2,1}\sim :\varphi^{m-1}:\,,~~{\rm with~~}\Delta=\{{3\over 2m(m+1)}\,,{m-2\over 2(m+1)}\,,{m+3\over 2m}\}\,,~
\label{minflows}
\fe
where we have included their conformal weights \footnote{In general, the chiral conformal weights of the Virasoro primary $\phi_{r,s}$ is $h_{r,s}={(r(m+1)- sm )^2-1\over 4m(m+1)}$.}. The corresponding bulk RG flows have been studied extensively by both the Truncated Conformal Space Approach (TCSA) and integrability (for the last two types in \eqref{minflows}) \cite{Zamolodchikov:1989hfa,Lassig:1990xy,Yurov:1991my,Zamolodchikov:1990xc,Smirnov:1991uw,Colomo:1991gw,Mussardo:1992uc}. Furthermore, the symmetries and the boundary conditions of these minimal models are well-known. The simple topological defects $\cL_{r,s}$ and boundaries $|r,s\ra$ are in one-to-one correspondence with the Virasoro primaries $\phi_{r,s}$ with $1\leq r\leq m+1,1\leq s\leq m$ subjected to the identification $\{r,s\}\leftrightarrow \{m+1-r,m-s\}$. In particular, the symmetry action on the local operators 
\ie 
\cL_{r_1,s_1} \phi_{r_2,s_2}={S_{{r_1,s_1};{r_2,s_2}}\over S_{{1,1};{r_s,s_2}}} \phi_{r_2,s_1}\,,
\label{symaction}
\fe
is given by the modular $S$-matrix,
\ie 
S_{{r_1,s_1};{r_2,s_2}}=2\sqrt{2\over m(m+1)}(-1)^{r_1 s_2+r_2s_1+1}\sin {\pi m s_1 s_2\over m+1}\sin {\pi (m+1) r_1r_2\over m}\,.
\label{mmSm}
\fe
This is compatible with the following fusion rule of the topological defects as a consequence of the Verlinde formula,
\ie 
\cL_{r_1,s_1}\circ \cL_{r_2,s_2}=\bigoplus_{r=1+|r_1-r_2|,r+r_1+r_2\in 2\mZ+1}^{\min(r_1+r_2-1,2m-1-r_1-r_2)} 
\bigoplus_{s=1+|s_1-s_2|,s+s_1+s_2\in 2\mZ+1}^{\min(s_1+s_2-1,2m+1-s_1-s_2)} 
\cL_{r,s}\,.
\label{mmfusion}
\fe
In particular $\cL_{1,m}$ generates the invertible $\mZ_2$ symmetry of $M_{m,m+1}$ and the charge of $\phi_{r,s}$ is $(-1)^{(m-1)r +ms +1}$.

The conformal boundaries are obtained by topological fusion with the identity brane $|1,1\ra$ whose $g$-function is determined by the S-matrix \eqref{mmSm},
\ie 
|r,s\ra = \cL_{r,s}|1,1\ra\,,\quad g_{|1,1\ra}=\sqrt{S_{1,1;1,1}}\,.
\fe
The multiplet structure of the boundaries under fusion with topological defects follow from \eqref{mmfusion}.

Below we will capitalize on these prior results to solve for the IR behaviors of these pinning flows following the strategy explained in Section~\ref{sec:2deg}.

Let us start with the simplest case $\cO=\phi_{2,2}$ (also known as the leading magnetization operator),  the bulk flow breaks all the symmetries and the bulk IR phase is trivially gapped, therefore from \eqref{symfact} we know the pinning defect is factorized into a simple boundary condition of $M_{m,m+1}$. Since this is the most relevant flow, we expect factorization via the boundary with the least $g$-function, which is the identity brane,
\ie 
 \cD_+(\phi_{2,2})=|1,1\ra\la 1,1|\,,\quad  \cD_-
 (\phi_{2,2})=|1,m\ra\la 1,m|\,.
\fe
The two flows are related by fusion with the $\mZ_2$ defect $\cL_{1,m}$ since $\phi_{2,2}$ is $\mZ_2$ odd.

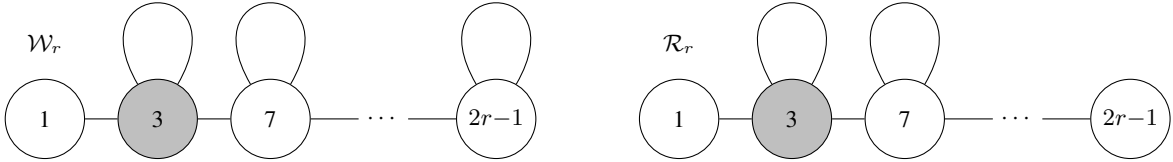
\begin{figure}[!htb]
\centering
\begin{minipage}[c]{.45\textwidth}
\centering
\begin{tikzpicture}[node distance=1.5cm, auto, >=stealth]
\node[circle, draw, minimum size=30pt, inner sep=1pt] (0) {1};
\node[circle, fill=gray!50, draw, minimum size=30pt, inner sep=1pt, right of=0] (1) {3};
\node[circle, draw, minimum size=30pt, right of=1] (2) {7};
\node[right of=2] (dots) {$\cdots$};
\node[circle, draw, minimum size=30pt, right of=dots] (3) {$2r{-}1$};
\draw[-] (0) -- (1);
\draw[-] (1) -- (2);
\draw[-] (2) -- (dots);
\draw[-] (dots) -- (3); 
\draw[-] (1) to[out=120, in=60, loop] (1);
\draw[-] (2) to[out=120, in=60, loop] (2);
\draw[-] (3) to[out=120, in=60, loop] (3);

\node[] at (0,1) {$\cW_r$};
\end{tikzpicture}
\end{minipage}
\begin{minipage}[c]{.45\textwidth}
\centering
\begin{tikzpicture}[node distance=1.5cm, auto, >=stealth]

\node[] at (0,1) {${\cal R}_r$};

\node[circle, draw, minimum size=30pt, inner sep=1pt] (0) {1};
\node[circle, fill=gray!50,draw, minimum size=30pt, inner sep=1pt, right of=0] (1) {3};
\node[circle, draw, minimum size=30pt, right of=1] (2) {7};
\node[right of=2] (dots) {$\cdots$};
\node[circle, draw, minimum size=30pt, right of=dots] (3) {$2r{-}1$};
\draw[-] (0) -- (1);
\draw[-] (1) -- (2);
\draw[-] (2) -- (dots);
\draw[-] (dots) -- (3); 
\draw[-] (1) to[out=120, in=60, loop] (1);
\draw[-] (2) to[out=120, in=60, loop] (2);
\end{tikzpicture}
\end{minipage}
    \caption{The fusion graphs for rank $r$ fusion categories $\cW_r$ and $\cR_r$. The shaded node is the generator of the graph. The $n$-th node corresponds to either $\cL_{1,2n-1}$ or $\cL_{2n-1,1}$ (see discussion around \eqref{smallrankWR}).}
    \label{fig:fgraph}
\end{figure}

 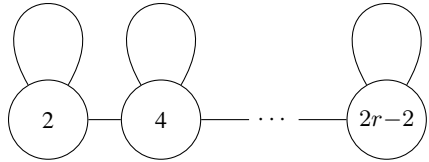
\begin{figure}
\centering
\begin{tikzpicture}[node distance=1.5cm, auto, >=stealth]

\node[circle,  draw, minimum size=30pt, inner sep=1pt, right of=0] (1) {2};
\node[circle, draw, minimum size=30pt, right of=1] (2) {4};
\node[right of=2] (dots) {$\cdots$};
\node[circle, draw, minimum size=30pt, right of=dots] (3) {$2r{-}2$};

\draw[-] (1) -- (2);
\draw[-] (2) -- (dots);
\draw[-] (dots) -- (3); 
\draw[-] (1) to[out=120, in=60, loop] (1);
\draw[-] (2) to[out=120, in=60, loop] (2);
\draw[-] (3) to[out=120, in=60, loop] (3);
\end{tikzpicture}
    \caption{The fusion graphs for the rank $r{-}1$ module category of  $\cR_r$. This module category can be represented in terms of the other topological defects in the minimal model, with the $n$-th node corresponding to either $\cL_{1,2n}$ or $\cL_{2n,1}$.}
    \label{fig:fgraphM}
\end{figure}

The pinning flows with $\cO=\phi_{2,1}$ or $\cO=\phi_{1,2}$ are more interesting because they preserve nontrivial symmetry subcategories. It is straightforward to verify that they are generated by topological defects of the type $\cL_{1,2i-1}$ or $\cL_{2i-1,1}$ respectively. The relevant fusion categories are presented in Figure~\ref{fig:fgraph} via their fusion graphs, where each node represents a simple object with the shaded node denoting the distinguished generator $Y$ of the fusion graph, and the edges between $X,Z$ keep track of the fusion channels in $X\circ Y \to Z$. For small ranks, these are well-known fusion categories (see $e.g.$ \cite{Chang:2018iay} and references therein)
\ie 
\cW_1=\cR_1={\rm Vec}\,,~ \cW_2={\rm Fib}\,,~  \cW_3={\rm Rep}(\widehat{so(3)}_5)\,,~ \cR_2={\rm Vec}_{\mZ_2}\,,~  \cR_3={\rm Rep}(S_3)\,.
\label{smallrankWR}
\fe
For $m$ even, the symmetry category preserved by $\phi_{2,1}$ is $\cW_{m\over 2}$ and it is $\cR_{m\over 2}$ for $\phi_{1,2}$. For $m$ odd, their roles are reversed, with $\cR_{m+1\over 2}$ for for $\phi_{1,2}$, and  $\cW_{m-1\over 2}$ for for $\phi_{1,2}$. This is intuitive since the adjacent minimal models $M_{m,m+1}$ and $M_{m-1,m}$ are related by an integrable RG flows triggered by $\phi_{1,3}$ \cite{Zamolodchikov:1989hfa}, and operators $\phi_{1,2}^{\rm UV}$ and  $\phi_{2,1}^{\rm IR}$ are related by the RG interface \cite{Gaiotto:2012np}.

The corresponding bulk flows in these cases are integrable \cite{Zamolodchikov:1989hfa} and the IR phases are known to be gapped and realized by TQFTs labeled by module categories for $\cW_r$ and $\cR_r$ depending on the sign of the deformation (see $e.g.$ \cite{Mussardo:2009ja} for a summary). The corresponding module categories are the regular module categories (from the module structure on the fusion categories themselves) and one rank $r{-}1$ module category over $\cR_r$ whose fusion graph is given in Figure~\ref{fig:fgraphM} (defined in a similar way as for the fusion graph of fusion categories and with the same generator as in Figure~\ref{fig:fgraph}).

Following the discussion in Section~\ref{sec:2deg}, we can then uniquely fix the factorization channels in \eqref{symfact} as follows, 
\ie 
 &\cD_+(\phi_{2,1})=\bigoplus_{i=1,3,\dots,m-1}|1,i\ra\la 1,i|\,,\quad \cD_-(\phi_{2,1})=\bigoplus_{i=2,4,\dots,m}|1,i\ra\la 1,i|\,,
 \\
 & \cD_+(\phi_{1,2})=\bigoplus_{i=1,3,\dots,m-1}|i,1\ra\la i,1|\,,\quad  \cD_-(\phi_{1,2})=\bigoplus_{i=2,4,\dots,m-2}|i,1\ra\la i,1|
\fe
for $m$ even, where the $\phi_{2,1}$ pinning defect factorizes through conformal boundaries in the regular module category of $\cW_{m\over 2}$ and the $\phi_{1,2}$ pinning defect factorizes through conformal boundaries in the regular module category and another rank $m-2\over 2$ module category of $\cR_{m\over 2}$ (see Figure~\ref{fig:fgraphM}). Note that the above are also consistent with the fact that $\phi_{2,1}$ anti-commutes with $\cL_{1,m}$ and $\phi_{1,2}$ anti-commutes with $\cL_{2,1}$ (see around \eqref{anticomfus}),
\ie 
\cL_{1,m}\circ \cD_\pm (\phi_{2,1})=  \cD_\mp (\phi_{2,1}) \circ \cL_{1,m}\,,\quad \cL_{2,1}\circ \cD_\pm (\phi_{1,2})=  \cD_\mp(\phi_{1,2}) \circ \cL_{2,1}\,.
\fe
The $m=4$ case also agrees with Table~\ref{tab:TIMflows} as expected.
Similarly for $m$ odd, the answers are
\ie 
 & \cD_+(\phi_{2,1})=\bigoplus_{i=1,3,\dots,m}|1,i\ra\la 1,i|\,,\quad  \cD_-(\phi_{2,1})=\bigoplus_{i=2,4,\dots,m-1}|1,i\ra\la 1,i|\,,
\\
&\cD_+(\phi_{1,2})=\bigoplus_{i=1,3,\dots,m-2}|i,1\ra\la i,1|\,,\quad \cD_-(\phi_{1,2})=\bigoplus_{i=2,4,\dots,m-1}|i,1\ra\la i,1|\,.
\label{modd}
\fe
which satisfy
\ie 
\cL_{2,1}\circ \cD_\pm (\phi_{2,1})=  \cD_\mp(\phi_{2,1}) \circ \cL_{2,1}\,,\quad \cL_{1,m}\circ \cD_\pm(\phi_{1,2})=  \cD_\mp(\phi_{1,2}) \circ \cL_{1,m}\,.
\fe

\subsection{Pinning flows in 3-state Potts model}

Pinning flows in non-diagonal Virasoro minimal models can be deduced from those in the diagonal theories by generalized gauging (orbifold) of non-invertible symmetries (see \cite{Diatlyk:2023fwf} for a review). In particular, the IR factorization property obviously persists if the symmetry being gauged commutes with the pinning flow.

Here we give one simple example of the pinning flow in the 3-state Potts model with $\cO=\phi_{2,1}$ of dimension $\Delta={4\over 5}$. The Potts model is related to the $M_{5,6}$ minimal model by gauging the $\mZ_2$ symmetry generated by the topological defect $\cL_{1,3}$. While $\phi_{2,2},\phi_{1,2}$ are $\mZ_2$ odd, $\phi_{2,1}$ is $\mZ_2$ even and thus the corresponding pinning defect $\cD_\pm (\phi_{2,1})$ survives the orbifold. Nonetheless, the reshuffling of the bulk operator spectrum under the orbifold changes the boundary states 
\cite{Affleck:1998nq,Fuchs:1998qn}, and the symmetries \cite{Petkova:2000ip,Chang:2018iay} and in turn, the IR behavior of $\cD_\pm (\phi_{2,1})_{\rm Potts}$. The eight conformal boundaries organize into a multiplet with respect to the bulk symmetries generated by the $\mZ_3$ topological defect $\omega$ satisfying $\omega^3=\id$, the Fib topological defect $W$, and the $\mZ_3$ Tambara-Yamagami duality defect $\cN$ satisfying $\cN^2=\id\oplus\omega\oplus\omega^2$ and $\omega \cN=\cN\omega=\cN$, 
\ie 
|1\ra={1\over \sqrt{2}}\left(|1,1\ra+|1,5\ra \right)\,,~|\omega^n\ra\equiv \omega^n|1\ra\,,~
|\cN\ra\equiv \cN|1\ra\,,~
|W\cN\ra\equiv W\cN|1\ra\,,~|W \omega^n\ra\equiv W|\omega^n\ra\,.
\fe
There is also a charge conjugation symmetry $C$ which extends the $\mZ_3$ to an $S_3$ symmetry and permutes the boundaries.
In the first equation above we give the explicit expression for the identity brane in the Potts model from those in $M(5,6)$.
As before, the $g$ functions of the general boundaries follow from topological fusion and the quantum dimensions of the topological defects. 

The symmetry subcategory that commutes with $\phi_{2,1}$ in the Potts model is ${\rm Vec}_{S_3}$.
The bulk RG flow is integrable and for the positive deformation, the IR phase is trivially gapped and for the negative deformation, there are three degenerate ground states \cite{Mussardo:1992uc,Lepori:2009ip}. Following the general discussion in Section~\ref{sec:symconstraint} and Section~\ref{sec:2deg}, the factorization channels of the pinning defect in the IR is fixed to be the following,
\ie 
\cD_+(\phi_{2,1})=|\cN\ra\la \cN|\,,\quad \cD_+(\phi_{2,1})=\bigoplus_{n=0}^2|\omega^n\ra\la \omega^n|\,,
\label{pottspin}
\fe
which are obviously related by $\cN \cD_\pm(\phi_{2,1})=\cD_\mp(\phi_{2,1})\cN$ (and have the same $g$-function)
since the operator $\phi_{2,1}$ anti-commutes with the defect $\cN$ \cite{Chang:2018iay} (also see around \eqref{anticom}). One can also check that \eqref{pottspin} follows from \eqref{modd} directly by the orbifold procedure \cite{Collier:2021ngi}.

\subsection{Pinning flows from nontrivial topological defects}

In the main text, we have briefly mentioned the generalization of our pinning flows to cases where the UV defect is a nontrivial topological defect $\cL_{\rm UV}$ and the operator $\cO$ is a general operator on $\cL_{\rm UV}$ (thus 
in general in a twisted sector when $\cL_{\rm UV}$ is non-invertible). Here we consider a few examples in the Ising CFT. Recall that the nontrivial topological defects here are the $\mZ_2$ generator $\eta$ and the duality defect $\cN$ (not to be confused with the $\mZ_3$ duality defect in the previous section). 

\begin{figure}[!htb]
\centering
\begin{minipage}[c]{.25\textwidth}
\centering
\begin{tikzpicture}[baseline={(current bounding box.center)}]

\draw (2,0) -- (2,2);
 
\node at (2.3,1.8) {$\cN$};
 
\draw[dashed] (2,1.0) -- (3,1.0);
 
\node at (3.5,1.0) {$\cO(x)$};
\filldraw (3,1.0) circle (1pt);
 
\node at (2.4,1.2) {$\eta$};
\end{tikzpicture}
\end{minipage}
\begin{minipage}[c]{.35\textwidth}
\centering
\begin{tikzpicture}[baseline={(current bounding box.center)}]
\draw (0,0) -- (0,2);

\node at (-1,1.0) {$\sigma(x)$};
\filldraw (-.5,1.0) circle (1pt);
 
\node at (0.3,1.8) {$\cN$};
 
\node at (1.,1) {$=$};
 
\draw (2,0) -- (2,2);
 
\node at (2.3,1.8) {$\cN$};
 
\draw[dashed] (2,1.0) -- (3,1.0);
 
\node at (3.5,1.0) {$\mu(x)$};
\filldraw (3,1.0) circle (1pt);
 
\node at (2.4,1.2) {$\eta$};
\end{tikzpicture}
\end{minipage}
\begin{minipage}[c]{.35\textwidth}
\centering
\begin{tikzpicture}[baseline={(current bounding box.center)}]
\draw (0,0) -- (0,2);
\draw (-1,0) -- (-1,2);

\node at (-.5,1.3) {$\sigma(x)$};
\filldraw (-.5,1.0) circle (1pt);
 
\node at (0.3,1.8) {$\cN$};
 \node at (-1.3,1.8) {$\cN$};
\node at (1.,1) {$=$};
 
\draw [thick](2,0) -- (2,2);
 
\node at (2.5,1.8) {$\id\oplus \eta$};
 
\draw[dashed] (2,1.0) -- (3,1.0);
 
\node at (3.5,1.0) {$\mu(x)$};
\filldraw (3,1.0) circle (1pt);
\node at (2.4,1.2) {$\eta$};
\end{tikzpicture}
\end{minipage}
    \caption{The first diagram defines the generalized pinning flows from $\cN$ with operator $\cO$ in the $\mZ_2$ twisted sector. The last two diagrams arise from
    topological moves in the Ising CFT relating different pinning flows. }
    \label{fig:isingpin}
\end{figure}
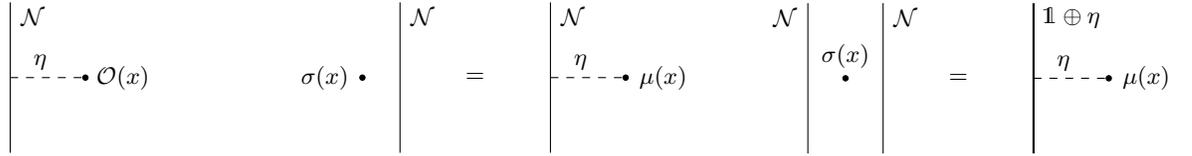

For simple $\cL_{\rm UV}$, the only interesting possibility is $\cN$ with primary operators in the $\mZ_2$ twisted sector (since $\eta$ is the only nontrivial fusion channel of $\cN$ with itself),
\ie 
\cH_\eta \ni \{\psi_{{1\over 2},0}\,,\tilde\psi_{0,{1\over 2}}\,,\m_{{1\over 16},{1\over 16}}\}
\,.
\label{Z2twistIsing}
\fe
We write the corresponding generalized pinning defect as $D^{\cN}_\pm(\cO)$ with $\cO$ among the list in \eqref{Z2twistIsing} with the orientation specified by the first diagram in Figure~\ref{fig:isingpin}. 

The flows triggered by $\psi,\tilde \psi$ (related by a parity flip) are studied in \cite{Runkel:2007wd,Kormos:2009sk}. Since these deforming operators are purely (anti)chiral, the corresponding defects are translation invariant along the entire flow, and thus must become topological in the IR \cite{Runkel:2007wd,Kormos:2009sk}. By $g$-theorem, the IR defect can either be $\eta$ or $\id$, and the choice is fixed by studying the fusion with conformal boundaries of the Ising CFT  \cite{Kormos:2009sk},
\ie 
D^\cN_+(\psi)=\eta\,,\quad D^\cN_-(\psi)=\id\,.
\label{Nflow}
\fe
Let us now consider the flow triggered by the disorder spin operator $\m$. From the left diagram in Figure~\ref{fig:isingpin}, we find that  this is related to the pinning flow from the trivial defect and then fused onto $\cN$,
\ie 
 \cD_\pm^\cN(\m) =\cD_\pm(\sigma)\cN\,,
\fe
from we which we conclude
\ie 
\cD_+^\cN(\m)=|f\ra \la +|\,,\quad \cD_-^\cN(\m)=|f\ra \la -|\,,
\fe
where $\{|+\ra,|-\ra=\eta|+\ra,|f\ra=\cN|+\ra\}$ are the three simple boundary conditions of the Ising CFT. 

There are also interesting flows when $\cL_{\rm UV}$ is not simple. For example, let us take $\cL_{\rm UV}=\id\oplus \eta$ and consider pinning defects $D^{\id\oplus\eta}_\pm(\cO)$ with $\cO$ from \eqref{Z2twistIsing} (defined in a similar way as in the first diagram in Figure~\ref{fig:isingpin}). In this case, clearly the IR behavior does not depend on the sign of the pinning field. Furthermore, from the last diagram in Figure~\ref{fig:isingpin}, we deduce immediately \footnote{See also \cite{Antinucci:2024izg} where the same flow was considered.}
\ie 
D^{\id\oplus\eta}_\pm(\m)=\cN D_\pm (\sigma) \cN=|f\ra\la f|\,.
\fe
Similarly, by \eqref{Nflow} and 
fusion with $\cN$, we also have
\ie 
D^{\id\oplus\eta}_\pm(\psi)=D^{\id\oplus\eta}_\pm(\tilde\psi)=\cN \,.
\fe

\section{Existence of self-adjoint extension of regularized $\hat{\cO}$}
\label{app:existence_of_extension}

In the main text, we have considered 
 the following sesquilinear  form defined for two states $|\phi_f\ra,|\phi_g\ra \in  \cH$ as defined in \eqref{genv} via the CFT three-point functions with the operator $\cO$ that defines the pinning defect,
\ie 
     \braket{\mathcal{\phi}_f |\hat{\cO}| \mathcal{\phi}_g} 
    =  \sum_{a,b} C_{ab \cO}\int_{\mR_+^d} \frac{d^d x d^d y d^{d-1}\vec z \,f_a(x) g^*_b(y) }{\left(|\vec x - \vec y|^2 + (x_d+y_d)^2\right)^{\frac{\Delta_a+\Delta_b - \Delta_\cO}{2}} \left( |\vec x - \vec z|^2 + x_d^2\right)^{\frac{\Delta_a+\Delta_\cO - \Delta_b}{2}} \left(|\vec y - \vec z|^2 + y_d^2\right)^{\frac{\Delta_b+\Delta_\cO - \Delta_a}{2}}}\,,\label{app:eq:matrix_elements}
\fe 
where for simplicity, we have included the contributions from scalar primaries in \eqref{genv}.
Although the RHS of \eqref{app:eq:matrix_elements} defines a Hermitian sesquilinear form densely on $\cH$, it does not guarantee the existence of a self-adjoint operator (which one naively would like to call $\hat \cO$) whose matrix elements coincide with the above on a dense subspace. This is because $\hat \cO$, as is defined in \eqref{genpindef}, is worse than unbounded, since it has a trivial domain in general. This is a consequence of non-integrable singularities in the $\cO\cO$ OPE (from identity operators and other light operators) that dominate the four-point function (in the $t$-channel) $\la \psi |\cO \cO |\psi\ra$ which governs the norm of a state $\hat \cO |\psi\ra$ for $|\psi\ra\in \cH$. 

To cure this divergent norm, we need to regularize $\hat \cO$. A naive guess would be to introduce an energy cutoff in the $s$-channel exchange (summing over all states created by $\hat \cO|\psi\ra$ with energy truncation). However, for our purpose, it's important to preserve conformal symmetry. Therefore we instead introduce a truncation in the conformal primaries which generate the subspace $\cH_{\Delta}\subset \cH$. Assuming discreteness in the CFT spectrum, this amounts to truncating the infinite sum over $s$-channel conformal blocks in $\la \psi |\cO \cO |\psi\ra$  to a finite sum. It's well-known that individual conformal blocks have softer divergences (logarithmic) in the $t$-channel OPE limit (see $e.g.$ \cite{Pappadopulo:2012jk,Mukhametzhanov:2018zja} for how such weaker divergences resum to produce the stronger $t$-channel singularity in the absence of truncation) and thus this should suffice to regularize $\hat \cO$ and we denote the corresponding regularized self-adjoint operator by $\hat \cO_\Delta$. Below we will establish the existence of such an operator, densely defined on $\cH_\Delta$, rigorously using Kato's representation theorems on quadratic forms \cite{kato2013perturbation} (see also 
\cite{reed1981functional,reed1975ii,simon2015comprehensive}).

On the truncated space $\cH_\Delta$, we have the following sesquilinear form
\ie 
 (\phi_f,\phi_g)
    =  \sum_{\Delta_{a,b} \leq \Delta} C_{ab \cO}\int_{\mR_+^d}\frac{d^d x d^d y d^{d-1}\vec z f_a(x) g^*_b(y) }{\left(|\vec x - \vec y|^2 + (x_d+y_d)^2\right)^{\frac{\Delta_a+\Delta_b - \Delta_\cO}{2}} \left( |\vec x - \vec z|^2 + x_d^2\right)^{\frac{\Delta_a+\Delta_\cO - \Delta_b}{2}} \left(|\vec y - \vec z|^2 + y_d^2\right)^{\frac{\Delta_b+\Delta_\cO - \Delta_a}{2}}}\,, 
    \label{truncateform}
\fe 
defined on the dense subspace $\Phi_{\Delta}\subset \cH_\Delta$ spanned by finite linear combination of the primary fields and their descendants from $\cH_\Delta$ (again only including scalar primary contributions in the sum over $a,b$ for simplicity).
We want to show that there is a self-adjoint operator $\hat \cO_\Delta$, such that it is densely defined on $\Phi_{\Delta}$ and
\ie 
\la \phi_f |\hat\cO_\Delta|\phi_g\ra=(\phi_f,\phi_g)\,.
\label{opform}
\fe 

First let us prove this in the case when $\cH_\Delta$ contains only one conformal family, which we denote as $\cH_\phi$, with scalar primary $\phi(x)$. The corresponding sesquilinear form is defined by the following matrix elements,
\ie 
     & \left.(\phi_f,\phi_g)\right|_{\cH_\phi}
    \equiv    B^{(1)}_\phi (x,y) B^{(2)}_\phi(x,y)\,,
    \label{phiformfactorize}
\fe 
where $B^{(1,2)}_\phi (x,y)$ define two sesquilinear forms separately by 
\ie 
      & B^{(1)}_\phi(x,y)  \equiv   \frac{1}{\left(|\vec x - \vec y|^2 + (x_d+y_d)^2\right)^{\frac{2\Delta_\phi - \Delta_\cO}{2}}} \,,\quad 
B^{(2)}_\phi(x,y) \equiv  \int d^{d-1} z \frac{1}{ \left( |\vec x - \vec z|^2 + x_d^2\right)^{\frac{\Delta_\cO}{2}} \left(|\vec y - \vec z|^2 + y_d^2\right)^{\frac{\Delta_\cO}{2}}}\,.
\label{B1B2}
\fe 
It is easy to see that $B^{(2)}_\phi (x,y)$ is  positive definite because 
\ie 
\int d^d x d^d y f(x) f^*(y)  \int d^{d-1} \vec z \frac{1}{ \left( |\vec x - \vec z|^2 + x_d^2\right)^{\frac{\Delta_\cO}{2}} \left(|\vec y - \vec z|^2 + y_d^2\right)^{\frac{\Delta_\cO}{2}}} 
= \int d^{d-1} \vec z \left| \int d^d x \frac{f(x)}{\left( |\vec x - \vec z|^2 + x_d^2\right)^{\frac{\Delta_\cO}{2}}} \right|^2\,.\label{eq:proof_of_semiboundness}
\fe 
Note that $B^{(1)}_\phi (x,y)$ is positive definite for any $\Delta_\phi > \Delta_\cO/2$
as a consequence of the following transformation
\ie
\int\limits_{\mathbb{R}_+^d} d^d x d^d y\frac{f(x) f^*(y)}{\left(|\vec x - \vec y|^2 + (x_d+y_d)^2\right)^{\frac{2\Delta_\phi - \Delta_\cO}{2}}}
 &   =  \int \limits_{\mathbb{R}_+^d} d^d x d^d y 
   \int_0^\infty dt \frac{t^{\frac{2\Delta_\phi - \Delta_\cO}{2} - 1} }{\Gamma\left(\frac{2\Delta_\phi - \Delta_\cO}{2}\right)}  f(x) f^*(y) e^{-t \left|\vec{x} - \vec y\right|^2 - t (x_d + y_d)^2} \\
  &=
   \sum_{n \geq 0 } {1\over n!} \int d^{d-1}\vec xd^{d-1}\vec y \int_0^\infty dt \frac{ t^{\frac{2\Delta_\phi - \Delta_\cO}{2} - 1} }{\Gamma\left(\frac{2\Delta_\phi - \Delta_\cO}{2}\right)}    e^{-t \left|\vec{x} - \vec{y}\right|^2} F(t,\vec x) F^*(t,\vec y)\,,
\label{posB1}
\fe
where 
\ie
F(t,\vec x)\equiv  \int_0^\infty  dx_d e^{-t x_d^2} x_d^n (2t)^\frac{n}{2} f(x_d, \vec x)\,.
\fe 
The positivity of \eqref{posB1} then follows from the Fourier transform, since the kernel
 $K_t(\vec{x},\vec{y}) \equiv  e^{-t\left|\vec{x} - \vec{y}\right|^2}$  satisfies,
\ie 
K_t(\vec{x},\vec{y}) = \int \frac{d^{d-1} \vec p}{(4\pi t)^\frac{d-1}{2}} e^{i \vec p\cdot (\vec x-\vec y) } K_t(\vec{p}), \quad K_t(\vec{p}) = e^{-\frac{p^2}{4t}} > 0\,.
\fe 
For $\Delta_\phi = \Delta_\cO/2$, the above argument does not directly apply. Nonetheless, since the kernel is simply $B^{(1)}_\phi \equiv 1$ in this case,  for any state $\ket{\phi_f} \in \Phi_\Delta$, which we can approximate using a primary operator $\phi(x)$ as
\ie 
     \ket{\phi_f } = \sum\limits^n_{i=1} c_i \phi(x_i) \ket{0}\,, 
\fe 
 the positivity is obvious. The above argument, combined with 
 the Schur product theorem \cite{Horn_Johnson_1985} for the Hadamard product in \eqref{phiformfactorize} then implies that the sesqulinear form $\braket{\phi(x) |\hat{\cO}_\phi| \mathcal{\phi}(y)}$ is positive definite for $\Delta_\phi \geq \Delta_\cO/2$. Given the discussion above \eqref{truncateform}, we expect this positivity to persist for more general $\Delta_\phi$ consistent with unitarity, though the sesqulinear  forms  in \eqref{B1B2} may not be separately positive.

Kato's (second) representation theorem \cite{kato2013perturbation,reed1981functional,simon2015comprehensive} states that a densely defined closed Hermitian form is uniquely represented by a self-adjoint operator. The Hermitian forms and quadratic forms are clearly in one-to-one correspondence. The closeness condition amounts to requiring the corresponding quadratic form $q(\cdot)$ to satisfy lower semicontinuity, namely, for any convergent sequence $(f_n)$ \footnote{
	To illustrate the importance of the closed-ness condition, let us give one simple example with Hilbert space $\cH=L^2(\mathbb{R})$ and its dense subspace of Schwarz functions denoted by $\cS$.  The densely-defined quadratic form $q(f) = f(0)^2$  for $f \in \mathcal{S}$ is clearly positive but not closed, because the sequence $f_n = e^{-n^2 x^2} \in \mathcal{S}$ goes to zero with respect to the norm on $\cH$ but $f_n(0) = 1$. This form has a Dirac delta function kernel and is well-known to be not representible by a densely defined self-adjoint operator.},
\ie 
\lim_{n\to \infty}||f_n -f||=0 \Rightarrow q(f)\leq \lim_{n \to \infty} \inf q(f_n)\,.
\fe

Since the form defined by \eqref{phiformfactorize} and \eqref{B1B2} is closed by construction and positive, by Kato's theorem, we confirm the existence of the unique self-adjoint operator  $\hat{\cO}_\phi$
whose domain is dense in $\cH_{\phi}$ and whose matrix elements coincide with the form as in \eqref{opform}.

Now let us consider the case where two conformal families $\left\{\phi_{1,2}\right\}$ are involved in the definition of the form \eqref{truncateform}. The generalization to cases involving more conformal families can be done analogously. For this purpose it suffices to consider the following state $|\phi_f\ra \equiv  c_1 |\phi_1(x)\ra + c_2 |\phi_2(y)\ra $. The corresponding sesquilinear form reads,
\ie 
 &(\phi_f,\phi_f) = c_{11\cO}\left|c_1\right|^2 \braket{\phi_1(x) | \hat{\cO}_{\phi_1} | \phi_1(x)} + c_{22\cO}\left|c_2\right|^2 \braket{\phi_2(y) | \hat{\cO}_{\phi_2} | \phi_2(y)} 
 \\
 &+
    \frac{c_{12\cO}}{\left(|\vec x - \vec y|^2 + (x_d+y_d)^2\right)^{\frac{\Delta_1 + \Delta_2 - \Delta_\cO}{2}}} \int d^{d-1}\vec z \left[
  \left|\frac{c_1}{\left((\vec{x} - \vec{z})^2 + x_d^2\right)^{\frac{\Delta_1 + \Delta - \Delta_2}{2}}}
        + \frac{c_2}{\left((\vec{x} - \vec{z})^2 + x_d^2\right)^{\frac{\Delta_1 + \Delta - \Delta_2}{2}}}\right|^2
  - \left(- \leftrightarrow +\right)
\right]\,.
\fe
Since the above can be written as a sum of semi-bounded closed quadratic forms and each can be represented by a unique self-adjoint operator by Kato's theorem, we have identified the self-adjoint operator $\hat \cO_\Delta$ acting on two conformal families. Now this approach can be extended to $\cH_\Delta$ which contains a finite number of conformal families.

\section{Establishing the trivial representation}
\label{sec:important_lemma}

Given a Gelfand triple $\Phi_\mathcal{D} \subset \mathcal{H} \subset \Phi'_\mathcal{D}$ associated with the CFT Hilbert space $\cH$, and let \(\ket{\vec x} \in \Phi'_\mathcal{D}\) be elements obeying the following actions of the conformal subalgebra,
\ie 
    P_i |\vec x\ra 
    =  i\partial_i \ket{\vec x}\,, \quad 
    D|\vec x\ra = x_i \partial_i \ket{\vec x}\,,\quad 
    K_i |\vec x\ra =  i (2 x_i x_j\partial_j - x_j x_j\partial_i) \ket{\vec x}\,, \quad 
    M_{ij} |\vec x\ra = i(x_i \partial_j - x_j\partial_i) \ket{\vec x}\,, \label{eq:conf_group_x}
\fe 
we prove below that these states, when represented in terms of bulk local operators, as in 
\ie 
    |\vec x\ra = \sum_a \int_{\mR_+^d} d^d y\, f_a(\vec x,y)\, \phi_a(y) |0\ra\,,
    \label{genxdecomp}
\fe 
for an orthonormal basis of primary operators $\phi_a(x)$, are linear combinations of Ishibashi states. In particular, the coefficient function $f_a(\vec x,y)$ is only nonzero if $\phi_a$ is a scalar primary, in which case, up to an overall constant,
\ie
f_a(\vec x,y)= y_d^{\Delta_a-d}\,.
\label{faIshi}
\fe

Without loss of generality, we can focus on a single primary operator $\phi_a$ in \eqref{genxdecomp} and thus will drop the subscript $a$ below. We present the details below for scalar $\phi_a$ only as the case of spinning operators is very analogous. 

From the action of $P_i$ in \eqref{eq:conf_group_x} and \eqref{genxdecomp}, we have 
\ie 
    P_i |\vec x\ra =  i\int d^d y  f(\vec x,y)\partial_i \phi(y) |0\ra 
    =  -i\int d^d y \,\partial^y_i f(\vec x,y)\phi(y)|0\ra 
    =   i\partial^x_i \int d^d y  f(\vec x,y)\phi(y)|0\ra \,,
\fe 
which implies 
\ie 
\int d^d y (\partial^x_i + \partial^y_i) f(\vec x,y) \phi(x) |0\ra = 0 ~\Rightarrow ~
f(\vec x,y)=g(\vec x-\vec y,y_d)\,,
\fe 
where we have used the linear independent of $\phi(x)$. In a similar way, from the actions of rotation $M_{ij}$ and dilation $D$, we can further constrain the coefficient function $f(\vec x, y)$ to be 
\ie 
f(\vec x,y)=y_d^{\Delta-d}h\left(|\vec x -\vec y |\over y_d\right)\,.
\fe
Finally, the action of the special conformal transformation $K_i$ in \eqref{eq:conf_group_x} requires,
\ie 
  \left(\left(2 x_i x_j \partial^x_j - x^2 \partial^x_i\right) +  \left(2 y_i y_j \partial^y_j - y^2 \partial^y_i\right)+2y_i ( y_d \partial_d^y+d-\Delta_a)- y_d^2\partial_i^y  \right)\left[y_d^{\Delta_a - d} h\left(|\vec x -\vec y |\over y_d\right) \right]  = 0\,,
\fe
which forces $h$ to be a constant. The desired result \eqref{faIshi} then follows.

\section{Proving the weak convergence to zero state by dilation}
\label{sec:alternative_proof}
Here we prove that for an arbitrary scalar state $|\psi\ra\in \cH$, the following weak limit is trivial,
\ie
\underset{b\to \infty}{\operatorname{w-lim}}\,U_b\ket{\psi} = 0\,,
\label{vanishingwlim}
\fe 
where $U_b=e^{b D}$ is the unitary operator for dilation. 
Without loss of generality, we can assume that 
\ie 
|\psi \ra = \int_{\mR^d_+} d^d x\, f(x) \phi(x)|0\ra\,,
\fe 
namely $|\psi\ra$ is constructed from a single conformal family for a scalar primary operator $\phi(x)$ of dimension $\Delta$. 
Shifting the quantization surface by $\D$ in the $x_d$ direction and then using reflection positivity together with the Cauchy-Schwarz inequality, we obtain
\ie 
 \forall\, x_d \geq \D \geq 0:\quad \left|\braket{\phi(x)| \psi }\right|^2 \leq \frac{\cA_\psi(2t)}{(2(x_d - \D))^{2\Delta}}\,,\quad \cA_\psi(\D) \equiv \braket{\psi| e^{ \D P_d} |\psi } \geq 0\,.
\fe 
 By setting $\D = x_d/2$, this produces the following upper bound on the overlap by the amplitude $\cA_\psi(x_d)$,
\ie
 \left|\braket{\phi(x)| \psi}\right| \leq \frac{\sqrt{\cA_\psi(x_d)}}{x_d^\Delta}
~~\Rightarrow~~
 \left|\braket{\phi(x)| U_b \psi}\right|  \leq \frac{\sqrt{\cA_\psi(e^b x_d)}}{x_d^\Delta}\,,
 \fe 
 and the second inequality follows from dilation symmetry. 
Hence to establish \eqref{vanishingwlim} it suffices to show that
\ie \lim_{\D\to \infty} \cA_\psi(\D) = 0\,,
\label{vanamp}
\fe
which we derive below. To this end, we note that, explicitly
\ie 
\cA_\psi(\D)= \int_{x_d,y_d\geq 0} d^dx\, d^dy\, \frac{f(\vec x,x_d)\, f^*(\vec y,y_d)}{\left(\left(x_d + y_d + \D\right)^2 + |\vec x - \vec y|^2\right)^{\Delta}} \,,
\fe
which can be bounded by splitting the integration regions as below,
\ie 
\cA_\psi(\D)\leq \int_{|x|{\rm or}|y|\geq M} d^dx\, d^dy\, \frac{f(\vec x,x_d)\, f^*(\vec y,y_d)}{\left(\left(x_d + y_d\right)^2 + |\vec x-\vec y|^2\right)^{\Delta}} + \D^{-2\Delta}\left|\int_{|x|\leq M} d^dx\, f(x_i, x_d)\right|^2\,.
\fe 
Since $|\psi|^2=\cA_\psi(0)$ is finite by assumption, the first contribution on the RHS above can be made arbitrarily small by choosing  sufficiently large $M$ and the second contribution can then be arbitrarily suppressed (for fixed M) by sufficiently large $\D$. In other words, for any $\epsilon>0$, we can choose $M_\epsilon$ and $\D_\epsilon$ such that $\cA_\psi(\D)<\epsilon$ for all $\D>\D_\epsilon$, thus we have proved \eqref{vanamp}.

\section{Pinning defects in the free field theories}
\label{app:free_field}

Pinning defects in free theories tend to have exotic ($e.g.$ run-away) behavior in the IR \cite{Cuomo:2021kfm,Cuomo:2022xgw,Cuomo:2023qvp}. Here we consider the case
of a codimension-one pinning defect in the free scalar field theory,
\ie 
    S = \int d^d{x} \,\frac12 \left(\partial_\mu \phi\right)^2, \quad \hat{\cO} = \int d^{d-1}\vec x\, \phi(x,x_d=0)\,,
\fe 
and elaborate on its non-factorizing IR behavior in our framework as described in the main text.

This defect already exposes a problem at the level of the matrix elements for $\hat \cO$ \eqref{app:eq:matrix_elements}. In fact, the following integral
\ie 
    \langle \phi^2(x)|\hat{\cO}| \phi(y)\rangle = 2\int d^{d-1}\vec z\, \frac{1}{(x-y)^{d-2}} \frac{1}{\left(|\vec x-\vec z|^2 + x_d^2\right)^{\frac{d-2}{2}}},
\fe 
suffers from IR divergence. This divergence is related to the presence of vacuum moduli in the free scalar field (which fix the boundary condition at infinity). One could resolve this issue by performing an appropriate subtraction of these contributions, but doing so would render the operator $\hat{\cO}$ non-self-adjoint. 

A possible way to circumvent this problem is to study the defect on the cylinder $\mathbb{R}_\tau \times {\rm S}^{d-1}$, where we define
\[
\hat{\cO} = \int dS^{d-1} \phi(\vec x,\tau = 0)\,.
\]
For explicitness let us focus on $3d$ (the extension to general $d$ is straightforward). By canonical quantization, 
\ie 
    \phi(\vec x,\tau) = \sum_{l=0}^\infty \sum\limits^l_{m=-l}  
 Y_{l,m}(\vec x)\frac{\hat{a}^\dagger_{m,l} e^{\omega_l \tau} + \hat{a}_{m,l} e^{-\omega_l \tau} }{\sqrt{2\omega_l}}\,, \quad \left[\hat{a}_{l,m}, \hat{a}^\dagger_{l',m'}\right] = \delta_{ll'} \delta_{mm'}\,, \quad \omega_l = l +\frac12\,,
\fe 
where $Y_{l,m}$ denotes the ${\rm S}^2$ spherical harmonics,
we have 
\ie 
\hat{\cO} = \hat{a}_0 + \hat{a}_0^\dagger\,,
\fe 
which commutes with all other operators $\hat{a}_{l, m}$ and whose spectral decomposition is given by
\ie 
\hat\cO =  \int d\lambda \prod_{l>0,|m|
	\leq l} d\alpha_{l,m} \, \lambda |\lambda, \alpha_{l,m} \ra\la\lambda, \alpha_{l,m} |\,,\quad \hat \cO |\lambda, \alpha_{l,m}\ra=\lambda |\lambda\,, \alpha_{l,m}\ra\,, \quad \hat{a}_{l,m} |\lambda, \alpha_{l,m} \ra =  \alpha_{l,m} |\lambda, \alpha_{l,m} \ra\,.
 \fe 
Thus, for any element $\ket{\psi}\in \mathcal{H}$, by completeness, we have
\begin{gather}
    \braket{\psi|\psi} = \int d\lambda\, \prod_{l>0,|m|
    \leq l} d\alpha_{l,m} \, \braket{\psi|\lambda,\alpha_{l,m}} \braket{\lambda, \alpha_{l,m}|\psi}.
\end{gather}
Since the measure of $\lambda$ is flat, we have that  the weak limit of $\ket{\lambda}$ is zero, 
\ie
\underset{\lambda\to\infty}{\operatorname{w-lim}}\,\ket{\lambda} = 0\,,
\fe 
which produces an ill-defined non-conformal limit (correlation functions with $\phi(x)$ diverge as $x_d\to \infty$) and the lack of IR factorization. Let us stress that, in comparison to the general discussion in \eqref{specdensity}, for this particular case, any choice of $\ket{e} \in \cH$ would never give us the flat measure (due to the flat moduli preserved by the defect). Consequently, we have to enlarge the range of integration in the spectral decomposition from finite range to infinite, which produces the above exotic behavior.

\section{The ill-defined defect ``one-point'' functions}
\label{app:ishibashioverlap}
A careful reader might note that, from the definition of Ishibashi (and defect) state via the weak limit (see around \eqref{weaklimstate}), some correlation functions are ill-defined in the presence of a defect, even for normalizable states. This is not surprising, since we know that Ishibashi states are non-normalizable, and by the Riesz representation theorem \cite{kolmogorov1957elements} we must conclude that correlation functions involving Ishibashi states cannot be defined for all states in the Hilbert space $\cH$. Here we give a simple example in the N-S quantization \cite{Rychkov:2016iqz} since this is our choice of frame in the main text. We will also illustrate the same phenomena in the more familiar radial quantization. 

Let us consider a scalar primary operator $\phi$ with scaling dimension $\Delta$, which satisfies the following two-point function and Ishibashi state condition
\ie 
    \braket{\phi(x) \phi(y)} = \frac{1}{(x-y)^{2\Delta}}\,,\quad \lla\phi|\phi(x)\rangle = \frac{1}{x_{d}^\Delta}\,.
\fe 
We define the following state
\ie 
   | F_\phi \ra = \sum^\infty_{n=2} \phi\left(x_i = 0, x_d =  \left(n \log n\right)^\frac{1}{\Delta} \right) |0\ra\,.
   \label{badstate}
\fe 
which is normalizable. Explicitly, the norm is given by a series with positive terms,
\ie
 \langle F_\phi | F_\phi\rangle =& \sum_{n, m = 2}^\infty \frac{1}{\left(  \left(n \log n\right)^\frac{1}{\Delta}  +  \left(m \log m\right)^\frac{1}{\Delta} \right)^{2\Delta}} \,,
 \label{eq:norm}
\fe 
and we can bound its partial sum as below,
\ie
&\sum_{n, m = 2}^N \frac{1}{\left(  \left(n \log n\right)^\frac{1}{\Delta}  +  \left(m \log m\right)^\frac{1}{\Delta} \right)^{2\Delta}} =\sum^N_{n=2} \sum^n_{k=2} \frac{2}{\left(  \left(n \log n\right)^\frac{1}{\Delta}  +  \left(k \log k\right)^\frac{1}{\Delta} \right)^{2\Delta}}    
 \\
 &<  \sum^N_{n=2} \frac{2 n}{n^2 \log^2n} 
<   \frac{1}{\log^2 2} + \int\limits^{N+1}_{2} \frac{2 dx}{x \log^2 x} =  \frac{1}{\log^2 2} + \frac{2}{\log 2} - \frac{2}{\log(N+1)} \,.
\fe 
Consequently the series absolutely converges, 
\ie
   \langle F_\phi | F_\phi\rangle =  \lim_{N\to\infty}\sum_{n, m = 2}^N \frac{1}{\left(  \left(n \log n\right)^\frac{1}{\Delta}  +  \left(m \log m\right)^\frac{1}{\Delta} \right)^{2\Delta}} <  \frac{1}{\log^2 2} + \frac{2}{\log 2}\,.
\fe 
However, the overlap of \eqref{badstate} with the Ishibashi state is 
\ie
   {}& \lla \phi| F_\phi \rangle 
   = \sum^\infty_{n=2} \frac{1}{\left((n \log n)^\frac{1}{\Delta}\right)^\Delta}
   = \sum^\infty_{n=2} \frac{1}{n \log n}\,,
\fe 
and does not converge since the partial sum is unbounded,
\ie 
 \sum^N_{n=2} \frac{1}{n \log n} \geq \sum^N_{n=2} \int\limits^{n+1}_n \frac{dx}{x\log x} =  \int\limits^{N+1}_ {2} \frac{dx}{x\log x} = \log \log \left(N + 1\right) - \log \log 2\,.
 \fe

For completeness, let us provide another example in the spherical frame using radial quantization.
Let us consider a primary scalar state $\ket{\phi}$ on $S^{d-1}$. In this case the Ishibashi state is \cite{Nakayama:2015mva,Diatlyk:2024qpr},
\ie 
    |{\phi}\rra  = \sum_{n \geq 0} \kappa_n^\phi (P^2)^n   \ket{\phi}\,,\quad   \kappa_n^\phi= {2^{-2n}
\over n! (\Delta + 1- {d\over 2})_n}   \,,
\fe 
where $(a)_n\equiv \Gamma(n+a)/\Gamma(a)$ is the Pochhammer symbol. We take the following state with norm
\ie 
    \ket{\hat{\phi}} = \sum_{n\geq 0} \frac{\alpha_n}{\sqrt{
    \cN_n}} (P^{2})^n\ket{\phi}\,,\quad \braket{\hat{\phi}|\hat{\phi}} = \sum^\infty_{n = 0} \left|\alpha_n\right|^2\,,
    \label{badstate2}
\fe 
where 
\ie 
\cN_n\equiv \la \phi| (K^2)^n (P^2)^n |\phi\ra =16^n n! \left(\frac{d}{2}\right)_n (\Delta )_n \left(-\frac{d}{2}+\Delta +1\right)_n\,.
\fe
On the other hand, the overlap with the Ishibashi state is 
    \ie 
\la \hat{\phi}|\phi \rra = \sum_{n=0}^\infty \alpha_n^* \kappa_n \sqrt{\cN_n} \,.
\label{badstateoverlap2}
\fe 
Now if we pick $\alpha_n = \frac{1}{n}$ in \eqref{badstate2}, the state $\ket{\hat{\phi}}$ is clearly normalizable. However the overlap \eqref{badstateoverlap2} diverges because
\ie 
    \kappa_n \sqrt{\cN_n} =  n^{\frac{d-2}{2}} \left(  \sqrt{\frac{ \Gamma \left(-\frac{d}{2}+\Delta +1\right)}{\Gamma \left(\frac{d}{2}\right)
   \Gamma (\Delta )}} + \cO\left(n^{-1}\right) \right)
\fe 
at large $n$. Note that for $2d$, one obtains the familiar result that $\kappa_n \sqrt{\cN_n} = 1$.

\end{document}